\documentclass[11pt]{article}
\pdfoutput=1
\usepackage{jheppub}
\usepackage{float, extarrows, tikz-cd}
\usepackage{graphicx}
\usepackage{slashed}
\usepackage{tabularx,ragged2e}
\usepackage{amssymb}
\usepackage{amsmath,amssymb}
\usepackage{slashed}
\usepackage{hyperref}
\usepackage{caption}
\usepackage{xcolor}
\usepackage{dsfont}
\usepackage{verbatim}
\usepackage{subfig}
\usepackage{mathtools,xcolor,ytableau,amsfonts,tikz}
\usepackage{graphicx}
\usepackage{physics}
\usepackage{amsthm}
\usepackage[normalem]{ulem}

\newcolumntype{C}{>{\Centering\arraybackslash}X}

\newcommand{\address}[1]{\vbox{\center\em#1}}

\newcommand{\bP}{\bar{P}}
\newcommand{\bh}{\bar{h}}

\newcommand{\adsthree}{\text{AdS}_3}
\newcommand{\mc}{\mathfrak{m}}
\newcommand{\sgn}{\text{ sgn}}
\ytableausetup{centertableaux,boxsize=.5em}

\usepackage{tikz}
\newcommand*{\boxcolor}{black}
\makeatletter
\renewcommand{\boxed}[1]{\textcolor{\boxcolor}{%
\tikz[baseline={([yshift=-1ex]current bounding box.center)}] \node [rectangle, minimum width=1ex,rounded corners,draw] {\normalcolor\m@th$\;\,\displaystyle#1\;\,$};}}
 \makeatother

\begin{document}

\thispagestyle{empty}

\begin{center}

{\LARGE \bf {Symmetries and spectral statistics\\ in chaotic conformal field theories}}
\end{center}

\bigskip \noindent

\bigskip

\begin{center}

Felix M.~Haehl,$^a$ Charles Marteau,$^b$ Wyatt Reeves,$^b$ and Moshe Rozali$^b$

\address{a) School of Mathematical Sciences and STAG Research Centre,\\ University of Southampton, SO17 1BJ, U.K.}

\address{b) Department of Physics and Astronomy, University of British Columbia,\\ Vancouver, V6T 1Z1, Canada}
\vspace{0.5in}
    
{\tt f.m.haehl@soton.ac.uk, wreeves@phas.ubc.ca,\\ cmarteau@phas.ubc.ca, rozali@phas.ubc.ca}

\bigskip

\vspace{1cm}

\end{center}

We discuss spectral correlations in coarse-grained chaotic two-dimensional CFTs with large central charge. We study a partition function describing the dense part of the spectrum of primary states in a way that disentangles the chaotic properties of the spectrum from those which are a consequence of Virasoro symmetry and modular invariance. We argue that random matrix universality in the near-extremal limit is an independent feature of each spin sector separately; this is a non-trivial statement because the exact spectrum is fully determined by only the spectrum of spin zero primaries and those of a single non-zero spin (``spectral determinacy''). We then describe an argument analogous to the one leading to Cardy's formula for the averaged density of states, but in our case applying it to spectral correlations: assuming statistical universalities in the near-extremal spectrum in all spin sectors, we find similar random matrix universality in a large spin regime far from extremality.

\newpage

\setcounter{tocdepth}{3}
{}
\vfill
\tableofcontents

\vspace{10pt}
\section{Introduction}

A great deal of progress on understanding quantum chaos in conformal field theories (CFTs) has been made in the last decade, particularly thanks to holography. Scrambling and random matrix-like behaviour discovered in black holes \cite{Shenker:2013pqa,Saad:2018bqo,Cotler:2020hgz,Cotler:2020ugk,Eberhardt:2022wlc} has shown the importance of understanding how chaos manifests in holographic CFTs. While many results for the consequences of chaos for {\it correlation functions} have been discovered (e.g., scrambling \cite{Roberts:2014ifa}, the eigenstate thermalization hypothesis \cite{Lashkari:2016vgj}, and OPE coefficients \cite{Collier:2019weq}), the more traditional characteristic of quantum chaos, i.e., random matrix-like statistics in the coarse-grained spectrum of the theory \cite{Wigner1955,Dyson:1962es,Mirlin:2000cla,DAlessio:2015qtq}, has not been fully explored. In this work we are interested in the interplay between random matrix-like statistics in the CFT spectrum and the many symmetries and constraints that all two-dimensional CFTs posses.

The precise spectrum of any CFT is subject to highly non-trivial constraints such as the bootstrap equations \cite{Poland:2018epd}. These constraints can be applied to the coarse-grained spectrum, i.e., the spectrum averaged over microcanonical windows. An example in two-dimensional CFTs is the Cardy formula \cite{Cardy:1986ie}; it gives a profile for the density of state at large energy that accurately describes an average over micro-canonical windows of the true (discrete) density of states. Its derivation simply amounts to reinterpreting the vacuum state on a torus in the modular transformed channel using modular invariance. It is thus an example of a symmetry (modular invariance) constraining some part of the spectrum based on knowledge of another part.

In this paper we focus on {\it correlations} in the spectrum of chaotic two-dimensional CFTs, such as those measured by the spectral two-point function, understood in the sense of microcanonical averaging. In quantum chaotic systems, such correlations obey the same statistics as an (appropriate) ensemble of random matrices. In \cite{Cotler:2020hgz, Cotler:2020ugk}, it was shown that  the {\it near-extremal correlations} of BTZ black hole microstates in $\adsthree$ gravity obey random matrix statistics; this suggests that a dual holographic two-dimensional CFT must have the same statistics described by such a wormhole amplitude. We wish to characterize in a precise sense the way in which two-dimensional CFTs can exhibit such a quantum chaotic spectrum. Since the CFTs of interest are quantum field theories and exhibit an enormous amount of symmetry, a careful analysis is needed to make the connection with (much simpler) random matrix theory.\footnote{ See also \cite{Dyer:2016pou,Benjamin:2018kre,Kudler-Flam:2019kxq,Mukhametzhanov:2020swe} for discussions and numerical results regarding the time scales associated with spectral form factors in CFTs, and in particular for rational theories where ergodicity can emerge in an approximate sense.}

\paragraph{Symmetries and chaos.} A signature of quantum chaos is the existence of universal correlations in the spectrum. When the quantum system enjoys a symmetry, much of the spectrum is fixed by the symmetry and the correlations due to chaos tend to wash away. Therefore, one has to remove all consequences of the symmetry in order to be able to see the universal statistics of the chaotic spectrum. For example, for a system with a single conserved charge, one has to isolate different charge sectors, obtaining a universal description of each charge sector separately, whereas different sectors are uncorrelated \cite{Kapec:2019ecr}. In exploring the concept of randomness and statistical universality in two-dimensional CFTs, our first task is to  take care of the large number of symmetries such systems enjoy: 
\begin{enumerate}
\item[$(1)$] Virasoro symmetry: treating the Virosoro symmetry is standard -- each state in the CFT is a primary of some conformal dimensions $(h,\bar{h})$ or a descendant of such a primary. The spectrum of descendants is fully determined by Virasoro symmetry. The full density of states is thus determined in terms of the density of primaries, which is the object we will focus on.\footnote{ The Virasoro symmetry is infinite dimensional, perhaps suggesting some degree of integrability. However, when we fix the symmetry completely, for large central charge the spectrum remains dense; the difference between the density of all states and that of Virasoro primaries only differs by replacing $c$ with $c-1$. Thus there is still an exponentially dense number of states that is not determined by symmetry. This is to be contrasted with integrable systems where fixing the charges determines the state uniquely. As explained below, the results of \cite{Cotler:2020ugk} provide an existence proof of CFTs with eigenvalue repulsion in the appropriate limit.}

\item[$(2)$] Each conformal field theory has a conserved momentum, which we refer to as spin. In the spirit of the above comments, to see chaos we focus on the density of states for a fixed spin sector.

\item[$(3)$] Modular invariance: the primary density of states for fixed spin is not yet free of symmetry constraints. There are large conformal transformations, the modular transformations, which correlate the density of primaries of different conformal dimensions. To disentangle randomness from consequences of modular invariance, we decompose the partition function in terms of objects that are already modular invariant, thus enabling us to discover aspects of the theory which can be (pseudo-)random. As is described in detail in the next section, this decomposition utilizes the recent work on harmonic analysis of two-dimensional CFTs \cite{Benjamin:2021ygh} (see also \cite{Collier:2022emf,Paul:2022piq}).
\end{enumerate}

Some of these concepts are illustrated in figure \ref{fig:spectrum}.
We will be focusing primarily on modular invariance in this paper, and the ways in which it does or does not constrain the appearance of chaos in two-dimensional CFTs. Note that our goal is not to derive chaotic properties of CFTs, but rather to assume them and explore their interplay with symmetries and modular invariance. Because the existence of random-matrix like statistics in CFTs is motivated by holography, we now comment on our perspective on its implications.

\paragraph{Comments on `ensemble averaging'.}
The presence of a nearly-continuous dense spectrum in gravity, together with the factorization problem \cite{Maldacena:2004rf,Saad:2018bqo} and related developments in two-dimensional JT gravity (e.g., \cite{Saad:2019pqd,Stanford:2019vob,Stanford:2020wkf,Blommaert:2019wfy}), has led to the suggestion that three-dimensional gravity might be dual to an ensemble of CFTs \cite{Cotler:2020ugk,Maloney:2020nni,Afkhami-Jeddi:2020ezh,Cotler:2020hgz}. We now comment on these suggestions and outline the approach we follow.

Quantum mechanical systems exhibit statistical universalities referred to as ``quantum chaos", which underlie the equilibrium properties of the system and its approach to equilibrium. For example, the high-energy spectrum is dense and can be approximated by a continuous density of states, whose low-point correlators are universal. For a single particle chaotic quantum system \cite{Bohigas:1983er}, one can check the predictions of quantum chaos by numerically diagonalizing the Hamiltonian of the system, binning the eigenvalues to create a continuous function, and calculating the density of states and its correlators. This is the sense in which we discuss chaotic properties, such as spectral statistics, of a single conformal field theory without resorting to any notion of ensembles of theories. Our main question is which aspects of the resulting density of states are universal, and how they are constrained or unconstrained by symmetry.

Quantum mechanical statistical universalities are argued for and described in an effective field theory language known as the Efetov sigma model \cite{Wegner,Efetov:1983xg} (for a recent review see \cite{Altland:2020ccq}), where they emerge as the low frequency description of the full system.\footnote{An interesting sigma-model to describe some statistical properties of  two-dimensional CFTs, namely of their operator product expansion (OPE) coefficients, was written down in \cite{Belin:2021ibv}, whereas here we are mainly interested in the operator spectrum. We hope that our results, perhaps in combination with the techniques developed in \cite{Belin:2021ibv}, can set the stage for a further understanding of the effective field theory of chaos in CFTs and its derivation from first principles.} Often there is a canonical model which is used to exemplify these universalities, and that model (most commonly random matrix theory) often involves built-in randomness. Since the quantum chaotic properties are universal, the canonical model can be chosen for reasons of convenience and any of its special features (e.g., ensemble averaging) have no special significance beyond simplifying calculations. 

\vspace{5pt}
\subsection{Summary of results}

\begin{figure}
    \centering
\includegraphics[width=.6\textwidth]{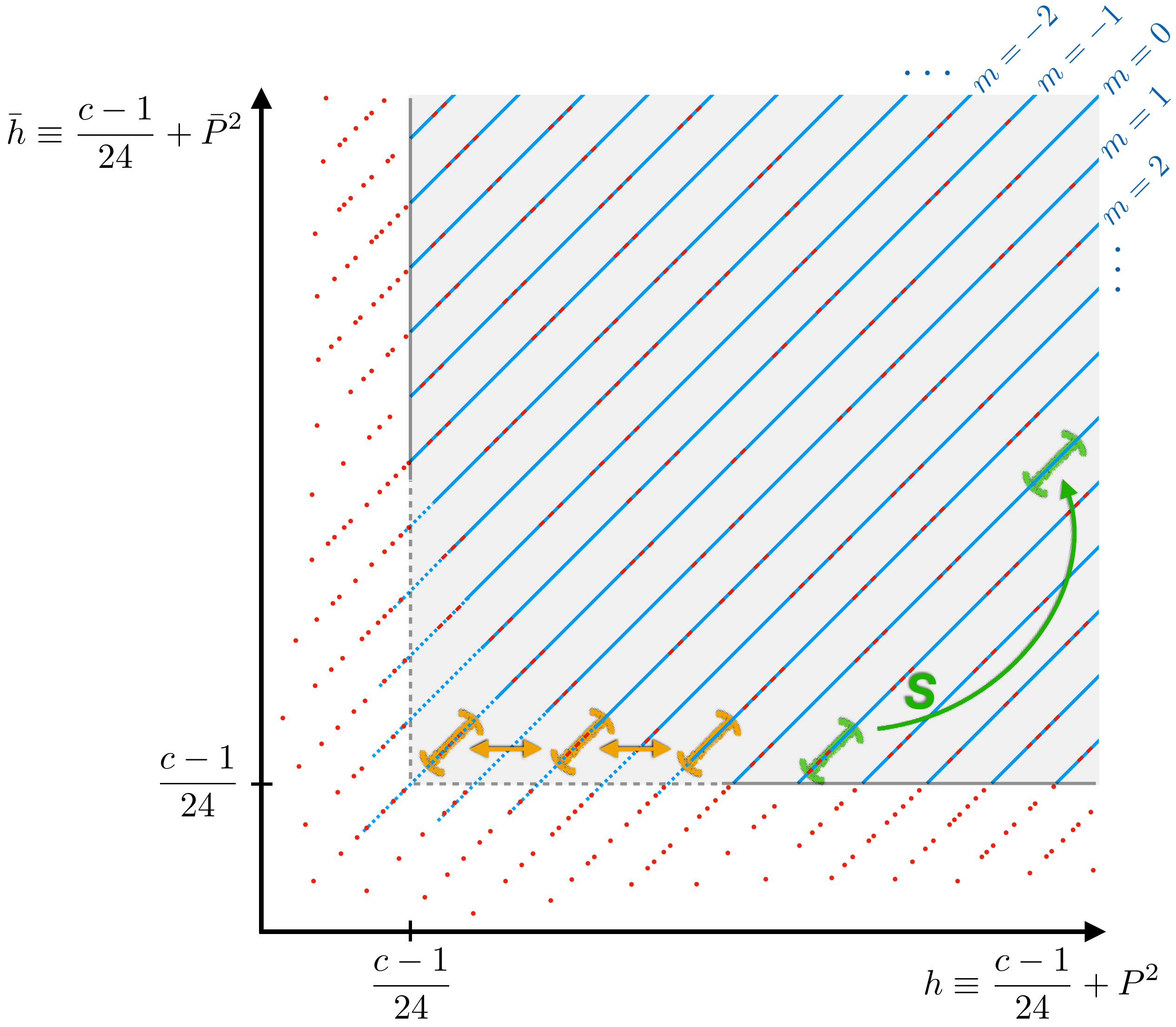}
    \caption{Cartoon of the spectrum of primary operators. The spectrum below the extremality bound (red dots outside the shaded region) is sparse and not chaotic. We remove it and all its modular images (red dots within the shaded region) from the primary partition function and focus on the statistics of the remaining dense spectrum after coarse-graining along fixed spin trajectories (blue). The first part of the paper (section \ref{sec:ramp}) demonstrates that chaos in different spin sectors near extremality (orange regions) contains independent information. The second part (section \ref{sec:cardy}) derives eigenvalue repulsion far from extremality by assuming random matrix statistics near extremality and  performing a modular S-transformation (green regions).
    }
    \label{fig:spectrum}
\end{figure}

Let us briefly summarize our two main findings regarding the interplay of modular invariance and quantum chaos. The first concerns the precise way in which modular invariance, when focusing on the near-extremal limit,  {\it doesn't} constrain quantum chaotic universalities (namely eigenvalue correlations) in the spectrum in different spin sectors and is hence consistent with more general expectations about the interplay of quantum chaos and symmetries. The second finding provides an example where modular invariance can lead to new types of universal behavior of the spectrum. See also figure \ref{fig:spectrum} for illustration.

\paragraph{Random matrix universality despite spectral determinacy.} It was shown in \cite{Benjamin:2021ygh}, based on the same techniques we will use below, just how constraining modular invariance is: {\it knowledge of the light ($h + \bar{h} \leq \frac{c-1}{12}$), the spin 0, and the spin $m$ spectrum for a single $m\geq 1$ completely determines the spectrum for all other spin sectors.} Following \cite{Benjamin:2021ygh}, we refer to this statement as spectral determinacy (see also \cite{Kaidi:2020ecu}). This statement concerns the exact spectrum. One of our goals is to elucidate the extent to which it holds for the coarse-grained spectrum and its statistical properties. We will give arguments that in the near-extremal limit spectral determinacy does not constrain the coarse-grained spectrum in the same way. In particular, we will argue for the following statement (under some assumptions): {\it in the long time/near-extremal limit, universal short-ranged eigenvalue repulsion of a quantum chaotic spectrum in any finite number of spin sectors is not sufficient to infer the same universality in all other spin sectors.} In other words, spectral determinacy is not constraining enough to prevent having random matrix statistics in each spin sector independently, nor powerful enough to infer random matrix statistics in every spin sector from a finite number of them.

\paragraph{Cardy-like formula for spectral correlations.} 
Assuming random matrix universality and therefore short-ranged eigenvalue repulsion for each spin sector in the near-extremal limit, we see how modular transformations relate it to interesting statements about eigenvalue correlations in the far-from-extremal regime. This is in the spirit of Cardy's derivation of the high energy behaviour of the (averaged) density of states, now applied to correlations in the spectrum. We show that the far-from-extremal spectrum also exhibits eigenvalue repulsion in a certain limit, demonstrating that random matrix universality applies in a larger part of the spectrum.

\vspace{5pt}
\subsection{Outline}

This paper is organized as follows: in section \ref{sec:modInv} we define the part of the spectrum which has all symmetry constraints removed, and review harmonic analysis and spectral determinacy. In section \ref{sec:ramp} we analyze quantum chaotic universality in spectral correlations across spin sectors in the near-extremal limit, exploring the consequences of spectral determinacy. Section \ref{sec:cardy} discusses Cardy-like constraints on the spectrum far from extremality, using assumptions about a universal near-extremal spectrum. We end with a discussion in section \ref{sec:discussion} and refer to appendices for some details and derivations.

\vspace{10pt}
\section{Modular invariant decomposition of the partition function}
\label{sec:modInv}

In this section we discuss the process of removing constraints of symmetry from the CFT spectrum, which results in the quantity of interest, the fluctuating part of the dense primary counting partition function. This is the quantity we use in the rest of the paper to search for statistical universalities.

\vspace{5pt}
\subsection{Dense primary counting partition function}

Let $Z(x,y)$ denote the partition function of the two-dimensional CFT under consideration, which is obtained by a path integral on a torus with modular parameter $\tau = x+iy$:
\begin{equation}
\label{eq:Zdef}
    Z(x,y)=\sum_{h, \bar{h}} e^{-2\pi y \left(h+\bar{h}-\frac{c}{12}\right)}\,e^{2\pi i x (h-\bar{h})}=\sum_{m} \int dE\; \rho^m(E) \,e^{-yE}\, e^{2\pi i x m} \,.
\end{equation}
Throughout, we will assume that the CFT only has Virasoro symmetry and no additional currents (and $c > 1$).
To enable statistical analysis we need a near-continuous density of states for sufficiently heavy operators, which is obtained in the limit of large central charge \cite{Hartman:2014oaa}. Following the above discussion, in order to strip the partition function from  consequences of symmetry we will now improve it in two steps (following \cite{Benjamin:2021ygh}):
\begin{enumerate}
    \item First, we wish to discard all constraints on the spectrum which are purely due to Virasoro symmetry.\footnote{ The fact that Virasoro descendants are not `ergodic' was also shown explicitly in \cite{Dyer:2016pou}.} This means, we want to consider a primary-counting partition function, which does not separately count Virasoro descendants. While this is achieved by multiplying the partition function by $\prod_{n\geq 1} |1-e^{2\pi i n \tau}|^2 \equiv e^{\frac{\pi y}{6}} |\eta(\tau)|^2$, this prescription alone is not modular invariant. We achieve the same in a modular invariant way if we define
     \begin{equation}
     \label{eq:ZPdef}
         Z_\text{P}(x,y) \equiv Z_0(x,y)^{-1} Z(x,y) \,,
     \end{equation}
     where $Z_0(x,y) = 1/(y^{1/2} |\eta(x+iy)|^2)$ is the partition function of a free massless non-compact boson. 
    \item Next, we split the primary states into an above-extremal, `non-censored' part (dimensions $h \text{ and }\bar{h} > \frac{c-1}{24}$) with a near-continuous dense spectrum, and a below-extremal, `censored' part ($h \text{ or }\bar{h} \leq \frac{c-1}{24}$) which is sparse \cite{Keller:2014xba}. We will hereafter refer to such primaries and their spectrum as dense and censored respectively. The censored spectrum is not dense and not expected to be chaotic (see, e.g., \cite{Schlenker:2022dyo}). We thus wish to focus only on the dense spectrum, but due to modular invariance the censored spectrum (particularly the vacuum) determines the coarse-grained behaviour of the dense spectrum.
    In order to focus on the pseudo-random aspects of the dense part of the spectrum we subtract off the `modular completion' of the sparse part of the spectrum, i.e., all modular images of the below-extremal primaries \cite{Keller:2014xba,Benjamin:2021ygh}. This defines:\footnote{In \cite{Benjamin:2021ygh} a similar object was defined using related but somewhat different rationale: if one averages (in some appropriate sense) over an ensemble of CFTs with fixed light spectrum, the subtracted part does not vary over the ensemble and therefore does not contribute to any correlations in the spectrum with respect to that ensemble average. Note that \cite{Benjamin:2021ygh} subtracts only light states with $h+\bar{h}\leq \frac{c-1}{12}$ and their modular images, as such states render the partition function non-normalizable. As we are interested in the chaotic part of the spectrum we subtract a larger part, i.e., the censored, sparse spectrum and its modular completion; spectral determinacy applies in both cases. We thank Eric Perlmutter for making this distinction clear to us.}
    \begin{equation}
    \label{eq:ZPfluctDef}
        \widetilde{Z}_\text{P}(x,y) \equiv Z_\text{P}(x,y) - \hat{Z}_C (x,y) \,,
    \end{equation}
    where the subscript $C$ stands for censored.
    As we wish to discuss correlations in the dense spectrum, the correct object to consider is $\widetilde{Z}_\text{P}(x,y)$. It computes the {\it deviations} from the universal dense average of the above-extremal, non-censored spectrum (e.g., due to the Cardy formula plus states dual to $SL(2,\mathbb{Z})$ black holes), with no censored primaries included.\footnote{Because there does not exist a unique definition of modular completion, different definitions can in some sense be seen as different definitions of coarse-graining the spectrum. We make the (weak) assumption that our modular completion features a continuous density of states.}

    The definition results in the fluctuating part of the dense primary counting partition function, namely forming a {\it normalizable function on the fundamental domain}, which is therefore suitable for harmonic analysis.
    \end{enumerate}
The modular invariant, fluctuating, dense partition function $\widetilde{Z}_\text{P}$ can now be analyzed using harmonic analysis on the fundamental domain. In particular, it is a modular invariant function of {\it slow growth} at the cusp $y\rightarrow\infty$ and can be decomposed into a basis of eigenfunctions of the Laplacian on the fundamental domain. These eigenfunctions are the real-analytic Eisenstein series $E_{\frac{1}{2}+i\alpha}(x+iy)$ with continuous label $\alpha \in \mathbb{R}$ and the Maass cusp forms $\nu_n(x+iy)$ with discrete labels $n \in \mathbb{Z}_+$ (in addition to a constant term $\nu_0 = \sqrt{3/\pi}$). 
The spin $m$ components of the Eisenstein series $E_s(x+iy) = \sum_{m\geq 0} \cos(2\pi m x) E_s^m(y)$ and of the Maass cusp forms $\nu_n(x+iy) = \sum_{m\geq 0}\cos(2\pi m x) \nu_n^m(y)$ are given by:
\begin{equation}
\begin{split}
E_{\frac{1}{2} + i \alpha}^{m=0}(y) &= \sqrt{y} \left[ y^{i\alpha} + \frac{\Lambda(i\alpha)}{\Lambda(- i \alpha)} \, y^{-i\alpha} \right] \,,\qquad\quad\;\; \nu_n^{m=0}(y) = 0 \,,\\
E_{\frac{1}{2} + i \alpha}^{m>0}(y) &=  \frac{4\,\sigma_{2i\alpha}(m)}{m^{i\alpha}\Lambda\left( -i\alpha\right)} \, \sqrt{y} K_{i\alpha}(2\pi m y) \,,\qquad\;\;
\nu_n^{m>0}(y) = a_m^{(n)} \, \sqrt{y} K_{iR_n} (2\pi m y) \,,
\end{split}\label{eq:EisMaassFourierComp}
\end{equation}
where $\Lambda(s) \equiv \pi^{-s} \Gamma(s) \zeta(2s)$ and $\sigma_z(n)$ is the divisor function.
The Maass cusp form coefficients $a_m^{(n)}$ and $R_n$ are real numbers that are known numerically, and they have no spin 0 component ($\nu_n^{m=0}\equiv 0$).\footnote{ For example: $\{R_n\} = \{13.780, \, 17.739, \, 19.423, \ldots\}$. We normalize $a_{m=1}^{(n)} = 1$, which implies for larger spins: $\{a_{m=2}^{(n)}\} =\{ 1.549,\, -0.765, \, -0.693,\ldots\}$, $\{a_{m=3}^{(n)}\} =\{ 0.247,\, -0.978, \, 1.562,\ldots\}$, etc.. See \cite{Benjamin:2021ygh} for some more data, \cite{lmfdb} for a comprehensive database, and \cite{Steil:1994ue} for an approximate expression describing the average distribution of $R_n$.}
Further properties of these functions in the context of spectral analysis on the fundamental domain are described in detail in \cite{Benjamin:2021ygh}.

This leads to a further decomposition of our basic object $\widetilde{Z}_\text{P}$ into: $(i)$ spin components labelled by $m$, and $(ii)$ the contribution to these spin components from the continuously and discretely labelled eigenfunctions of the Laplacian on the fundamental domain:
\begin{equation}
\label{eq:ZPtildeDef}
\begin{split}
 \widetilde{Z}_\text{P}(x,y) &= \widetilde{Z}^{m=0}_\text{P}(y) + 2\sum_{m\geq 1} \cos(2\pi m x) \widetilde{Z}^m_\text{P}(y) \,,\\
  \widetilde{Z}^m_\text{P}(y) &=  \widetilde{Z}^m_{\text{P,cont.}}(y) +  \widetilde{Z}^m_\text{P,disc.}(y)\,,
\end{split}
\end{equation}
where the decomposition into Eisenstein series and Maass cusp forms will be written as:
\begin{equation}
\label{eq:ZmPtildeDecom}
\widetilde{Z}^m_\text{P,cont.}(y) = \frac{1}{4\pi} \int_{-\infty}^\infty d\alpha  \,  z_{\frac{1}{2} + i\alpha} \,   E_{\frac{1}{2} + i \alpha}^m(y) \,,\qquad
  \widetilde{Z}^m_\text{P,disc.}(y)= \sum_{n\geq 0} \,  z_n \,   \nu_n^m(y)\,.
\end{equation}
In this section we assume a convention where $m\geq 0$ and we are focusing here on `even' (in $x$) Maass cusp forms.\footnote{In other words, the negative spin spectrum is fixed as $\widetilde{Z}^{-m}_\text{P,cont.}(y)=\widetilde{Z}^m_\text{P,cont.}(y)$.} More generally there can also be contributions proportional to $\sin(2\pi mx)$, but for simplicity we assume that the CFTs under consideration do not have such terms. The generalization is straightforward.

The absence of a spin 0 component in the Maass cusp forms results in a different large $y$ behaviour of the two pieces; the Eisenstein series grow like $\sqrt{y}$ as $y\rightarrow \infty$ (from the spin 0 component), while the Maass cusp forms decay like $e^{-2\pi y}$ (since all spin $m\geq 1$ components decay exponentially as $e^{-2\pi m y}$).

This decomposition leads to a change of perspective: the density of states in any given spin sector (or the partition function) is an object which may exhibit familiar signatures of chaos such as eigenvalue statistics, but it is the coefficients $z_{\frac{1}{2} + i\alpha}$ and $z_n$ which encode the spectrum in a modular invariant way. These coefficients for different values of $\alpha$ and $n$ are unrelated by symmetries and hence most natural to quantify chaos. The interplay of these  objects is the main tool we use below. 

The overlap coefficients in the decomposition \eqref{eq:ZmPtildeDecom} are given by:
\begin{equation}
\begin{split}
 z_{\frac{1}{2} + i\alpha} &\equiv \big( \widetilde{Z}_\text{P},\, E_{\frac{1}{2}+i\alpha} \big)  = \int_0^\infty dy' \, (y')^{-\frac{3}{2} -i\alpha} \, \widetilde{Z}_\text{P}^{m=0}(y')  \,,\\
 z_n &= \frac{(\widetilde{Z}_\text{P},\nu_n)}{(\nu_n,\nu_n)} \,.
\end{split}\label{eq:overlap-coeff}
\end{equation}
where the inner product used in this expression is defined in appendix \ref{app:notation}. We have used the `unfolding trick' to write the Eisenstein series overlap in terms of the spin 0 component of the partition function. This is one aspect of spectral determinacy: the spin 0 spectrum is sufficient to determine all continuously labelled coefficients $z_{\frac{1}{2}+i\alpha}$.\\

\vspace{5pt}
\subsection{Explicit spectral determinacy for nonzero spins}
\label{sec:xibasis}
The modular invariant decomposition of the partition function into (continuous and discrete) eigenfunctions of the Laplacian implies relations between different spin partition functions, as they all can be written in terms of a joint set of coefficients $z_{\frac{1}{2} + i\alpha}$ and $z_n$. These relations can be presented in different ways; here we shall now introduce a representation in terms of a variable $\xi$, which we will see plays a role similar to energy. This will be particularly useful to make the statement of spectral determinacy more explicit, as is shown presently, as well as discuss the relation between spectral statistics in different spin sector in the next section. 

In order to motivate the discussion below, recall that we are interested in the near-extremal and long time limit, in other words the behaviour near the cusp, $y \rightarrow \infty$. In that limit the spin $m>0$ partition function receives equal contribution from each of the basis functions, as their behaviour in that limit is identical, i.e., $\sqrt{y} K_{i\alpha}(2\pi m y) \sim \frac{1}{2\sqrt{m}}\,e^{-2\pi m y}$ for large $y$. In other words the contribution is expected to be localized in some conjugate variable, which we now construct.

Let us consider the following representation of the Bessel functions (for $m>0$):
\begin{equation}
    K_{i\alpha} (2\pi m y) = \frac{1}{2}\int_{-\infty}^\infty d\xi \; e^{-2\pi m y \cosh \xi} \, \cos(\alpha \xi) \label{eq:xi-definition}
\end{equation}
and the corresponding representation of the spin $m>0$ components of the partition function:
\begin{equation}
\label{eq:Zmfluct}
\begin{split}
    \widetilde{Z}^m_\text{P,cont.}(y)&= \frac{1}{\pi} \int_{-\infty}^\infty d\alpha \; z_{\frac{1}{2}+i\alpha} \, \frac{\sigma_{2i\alpha}(m)}{m^{i\alpha}\Lambda(-i\alpha)} \, \sqrt{y} K_{i\alpha}(2\pi m y) \\
    &= \frac{\sqrt{y}}{2\pi}\int_{-\infty}^\infty d\xi \,  {\cal Z}^m_\text{P,cont.}(\xi)\, e^{-2\pi m y \cosh \xi}  
\end{split}
\end{equation}
where the information about the coefficients $z_{\frac{1}{2}+i\alpha}$ is now encoded in 
\begin{equation}
\begin{split}
    {\cal Z}^m_\text{P,cont.}(\xi) &\equiv \int_{-\infty}^\infty d\alpha \,\cos(\alpha \xi)\, \frac{\sigma_{2i\alpha}(m)}{m^{i\alpha}\Lambda(-i\alpha)} \, \,z_{\frac{1}{2}+i\alpha} 
    \,,
\end{split}\label{eq:Zmpdisc}
\end{equation}
which is even in $\xi$.
Similar equations can be written for the Maass cusp forms. To aid readability, we give these in appendix \ref{app:maass} and focus here on the Eisenstein series.
Inverting \eqref{eq:Zmpdisc}, we find
\begin{equation}
\begin{split}
    \frac{\sigma_{2i\alpha}(m)}{m^{i\alpha}\Lambda(-i\alpha)} \, \,z_{\frac{1}{2}+i\alpha} &= \frac{1}{2\pi}\int_{-\infty}^\infty d\xi \, \cos(\alpha\xi) \, {\cal Z}^m_\text{P,cont.}(\xi) 
     \,.
\end{split}
\end{equation}
To summarize, we can present the (continuous) part of the spectrum in terms of three equivalent quantities: $(i)$ the partition function $\widetilde{Z}^m_{\text{P,cont.}}(y)$, $(ii)$  the transformed partition function ${\cal Z}^m_\text{P,cont.}(\xi)$, and $(iii)$ the coefficients $z_{\frac{1}{2}+i\alpha}$.

\paragraph{Spectral determinacy ($m>0$).}
Using the above representation, we can express the spin $m$ partition function (in $\xi$-space) as an integral transform of the spin $m'$ partition function:
\begin{equation}
\label{eq:calZcalZtrf}
\boxed{
    {\cal Z}^{m}_\text{P,cont.}(\xi) = \int_{-\infty}^\infty d\xi' \, {\cal K}_\text{cont.}^{m,m'}(\xi,\xi') \, {\cal Z}^{m'}_\text{P,cont.}(\xi') 
}
\end{equation}
with the following universal kernels:  
\begin{equation}
\label{eq:K-kernel}
    \begin{split}
    {\cal K}_\text{cont.}^{m,m'}(\xi,\xi') &= \frac{1}{2\pi}\int_{-\infty}^\infty d\alpha\, \cos(\alpha\xi) \cos(\alpha\xi')\, \frac{\sigma_{2i\alpha}(m)m'^{\,i\alpha}}{\sigma_{2i\alpha}(m')m^{i\alpha}} \,.
    \end{split}
\end{equation}
An analogous relation exists for the discrete part of the spectrum, see \eqref{eq:SpecDetMaass}.
Eq.\ \eqref{eq:calZcalZtrf} is the statement of spectral determinacy: ${\cal Z}^{m'}_\text{P,cont.}(\xi)$ for some $m'>0$ determines ${\cal Z}^{m}_\text{P,cont.}(\xi)$ for all other spins $m>0$. The transformation is also one-to-one and onto, as we can compose the kernels according to ${\cal K}^{m,m'} \circ {\cal K}^{m',m''}={\cal K}^{m,m''}$. Thus, any limited information in one spin sector (for example, knowing only the spectrum in spin $m$ for $\xi \in [-\xi_0,\xi_0]$) is equivalent to limited information in any other spin sector in a way specified by the above.

In summary, we have found that knowledge of the partition function in a single (non-zero) spin sector, and the decomposition of that partition function into continuous and discrete part, suffices to recover the partition function in any other spin sector. 

\vspace{5pt}
\section{Random matrix universality across spin sectors}
\label{sec:ramp}

In this section, we study the statistics of the chaotic spectrum and specifically two-point correlations in the limit of nearby energy levels. Due to spectral determinacy the following is obvious: {\it if} random matrix universality is exhibited by some part of the spin $m$ spectrum, we can infer this fact from {\it full} knowledge of the spectrum at spins 0 and a single $m'>0$ alone. Naively, this seems to be in tension with the intuition that different spin sectors should independently exhibit quantum chaos. The goal of this section is to show that there is no inconsistency, thanks to the fact that random matrix universality a priori concerns the near-extremal part of the spectrum, which is not subject to spectral determinacy in isolation.\\

In the following discussion we focus for simplicity on the part of the spectrum described by the Eisenstein series, which is more amenable to analytical arguments. The discrete $SL(2,\mathbb{Z})$ spectrum can {\it a priori} also be important for the physics we investigate, i.e., the statistical universalities we will discuss could be encoded fully or partially in the spectrum of cusp forms.\footnote{ We thank Scott Collier for helpful comments, which clarified this point for us.} We return to this point in the discussion section, and in \cite{toappear} we will show that all our results in the sector of Eisenstein series have an analog for the Maass cusp forms.

\subsection{Random matrix universality at fixed spin}

We begin by stating our assumptions and how they are motivated. One expects that the fixed-spin modular invariant primary partition function of dense states, $\widetilde{Z}^{m}_\text{P}(y)$, exhibits random matrix universality in the near-extremal limit. This universality is a statement about two-point correlations in the spectrum, so it depends on two copies of the partition function evaluated at $y_1$ and $y_2$. The near-extremal limit then consists of taking not only $y_i \rightarrow \infty$, but additionally holding $\frac{y_1}{y_2}$ fixed. Universal random matrix behavior in this regime corresponds to the following form of correlations:
\begin{equation}
\label{eq:ZPZPuniv}
 {\langle} \widetilde{Z}^{m_1}_\text{P}(y_1) \, \widetilde{Z}^{m_2}_\text{P}(y_2) {\rangle}_\text{ramp} = \frac{1}{2\pi} \, \frac{y_1 y_2}{y_1+y_2} \, \delta_{m_1m_2}\, e^{-2\pi (m_1y_1+m_2y_2)} + \ldots  \qquad \Big( y_i \gg 1, \; \frac{y_1}{y_2} = \text{fixed}\Big)\,.
\end{equation}
Note that all two-point functions in this paper refer to connected correlators. We will suppress a corresponding subscript throughout.
The subscript `ramp' refers to the fact that the spectral form factor, defined by the analytic continuation $y_1 = \beta + iT$, $y_2 = \beta - iT$, exhibits a specific $T$-dependence for large times $T$:
\begin{equation}
\label{eq:SFF0}
{\langle} \widetilde{Z}^{m_1}_\text{P}(\beta+iT) \, \widetilde{Z}^{m_2}_\text{P}(\beta-iT) {\rangle}_\text{ramp} \sim   \frac{T^2}{4\pi\beta} \, \delta_{m_1m_2}\, e^{-4\pi m_1 \beta  }+\ldots \qquad \left( T \gg \beta \gg 1 \right)\,.
\end{equation}
Note that the partition function used for harmonic analysis, defined in \eqref{eq:ZPdef}, involves multiplication by the free boson partition function $Z_0(x,y)$, which not only removes Virasoro descendent states, but also changes the growth at large $y$ by a factor $\sqrt{y}$ for each partition function and shifts the ground state energy by $\frac{\pi}{6}$. 
In order to recover the familiar form of the spectral form factor we remove these spurious factors when studying the large $y$ limit, and consider instead
\begin{equation}
\label{eq:ZPZPuniv2}
\;\; \frac{e^{\frac{\pi}{6}(y_1+y_2)}}{\sqrt{y_1y_2}}{\langle} \widetilde{Z}^{m_1}_\text{P}(y_1) \, \widetilde{Z}^{m_2}_\text{P}(y_2) {\rangle}_\text{ramp} = \frac{1}{2\pi} \, \frac{\sqrt{y_1 y_2}}{y_1+y_2} \, \delta_{m_1m_2}\, e^{-y_1 E_{m_1} -y_2 E_{m_2}} + \ldots  \quad \Big( y_i \gg 1, \; \frac{y_1}{y_2} = \text{fix}\Big)\,.\;\;
\end{equation}
where $E_{m} = 2\pi \left(m-\frac{1}{12}\right)$ is the ``ground state energy'' for the spin $m$ dense (BTZ black hole) spectrum. After analytic continuation to $y_{1,2}=\beta \pm iT$, this corresponds to removing a factor of $T$ from the spectral form factor \eqref{eq:SFF0} and shifting the ground state energy, thus recovering the well-known linear `ramp' (called like this and discussed in the context of black hole physics for the first time in \cite{Cotler:2016fpe}).

The expression \eqref{eq:ZPZPuniv} was argued for in holography \cite{Cotler:2020ugk,Cotler:2020hgz,Eberhardt:2022wlc} as it is the result of evaluating a two-boundary wormhole amplitude, which gives the leading gravitational contribution to the spectral form factor (this is also motivated by the black hole information problem \cite{Maldacena:2001kr,Dyer:2016pou}). Random matrix behavior in holographic two-dimensional CFTs near extremality is also expected based on the fact that this limit is described by a reduction to the Schwarzian theory \cite{Maxfield:2019hdt} (see also \cite{Mertens:2017mtv}): while the Schwarzian itself in the naive saddle point approximation is not sufficient to find a ramp in the spectral form factor, it is well understood that an alternative `wormhole' saddle contribution to the two-disk amplitude does reproduce it \cite{Cotler:2016fpe,Saad:2018bqo}. 
Finally, this is, of course, simply the expected behavior for the symmetry-unconstrained parts of the spectrum of generic chaotic quantum systems -- an expectation that is borne out in a vast number of different systems and hence very universal. 

To make the connection with traditional presentations, we change variables to energy eigenvalues (for fixed spin), where the above behavior translates into the universal form of eigenvalue repulsion for near-extremal energies:\footnote{We assume w.l.o.g.\ $E_1>E_2$.}
\begin{equation}
\begin{split}
    &\langle \widetilde{\rho}_\text{P}^{\,m_1}(E_1) \widetilde{\rho}_\text{P}^{\,m_2}(E_2)\rangle_{\text{ramp}} 
=  \int_{\epsilon-i\infty}^{\epsilon+i\infty} d y_1 dy_2 \left[  \frac{e^{\frac{\pi}{6}(y_1+y_2)}}{\sqrt{y_1y_2}}{\langle} \widetilde{Z}^{m_1}_\text{P}(y_1) \, \widetilde{Z}^{m_2}_\text{P}(y_2) {\rangle}_\text{ramp} \right] e^{y_1 E_1 +y_2 E_2} \\
    &\qquad\qquad\approx -\frac{1}{(2\pi)^2}\frac{E_1-E_{m_1}+ E_2-E_{m_2}}{\sqrt{E_1-E_{m_1}}\sqrt{E_2-E_{m_2}}\,(E_1-E_2)^2}\,\delta_{m_1m_2}  \qquad (0<E_i-E_{m_i} \ll 1) \\
    &\qquad\qquad \rightarrow -\frac{1}{2\pi^2 \omega^2}\,\delta_{m_1m_2} \qquad ( \omega \equiv E_1-E_2\ll E_i-E_{m_i})
\end{split}\label{eq:ramp-energy}
\end{equation}
where $\widetilde{\rho}^{\,m}_\text{P}(E)$ is the density of spin $m$ dense primary states in the symmetry-unconstrained part of the spectrum described by $\widetilde{Z}_\text{P}^m$. We refer to the limit taken in the second line as the {\it near-extremal limit}. In the last line we take additionally the small $\omega$ limit required to recover the familiar form of universal eigenvalue repulsion.

In the rest of this section, we study how information about the  universal form of the spectral form factor is encoded in the spectral decomposition of the partition function. 

\subsection{Independence of the ramp in each spin sector}
\label{sec:ramp-independence}

In this section, we present an analytic argument that demonstrates the independence of the ramp in each spin sector. By utilizing the definition of $\widetilde{Z}^{m}_\text{P,cont.}(y)$, \eqref{eq:Zmpdisc}, and the kernels \eqref{eq:K-kernel}, we will show the following: {\it assuming that any finite collection of spin sectors exhibits a ramp near extremality is never sufficient to conclude the existence of a ramp in all other spin sectors.}

First, we study how the ramp in the usual representation ($y$ variables) is encoded in the data amenable to spectral determinacy, $z_{\frac{1}{2}+\alpha}$ and ${\cal Z}^m_\text{P,cont.}(\xi)$. We will show that the spin $0$ ramp is equivalent to a particular large $\alpha_i$ behaviour of $\big\langle z_{\frac{1}{2}+\alpha_1}z_{\frac{1}{2}+\alpha_2}\big\rangle$. Similarly, ramps in spin sectors $m>0$ are equivalent to a particular small $\xi_i$ behaviour of $\big\langle {\cal Z}^m_\text{P,cont.}(\xi_1) {\cal Z}^m_\text{P,cont.}(\xi_2) \big\rangle$. 

We then use spectral determinacy to analyze how this data is interrelated. We will show that the presence of a ramp in the spin $m$ sector cannot be deduced from a ramp in any other spin sector, and in fact no finite collection of spin sectors is sufficient to find a ramp in all other spin sectors. 

\paragraph{Spin 0 ramp in $\alpha$ variables:} We begin by analyzing the spin 0 spectrum. We express $\widetilde{Z}_\text{P,cont.}^0(y)$ in terms of $z_{\frac{1}{2}+i\alpha}$:
\begin{equation}
  \widetilde{Z}^0_\text{P,cont.}(y) 
  = \frac{\sqrt{y}}{4\pi} \int_{\mathbb{R}} d\alpha \, z_{\frac{1}{2}+i\alpha}\, \left[ y^{i\alpha} + \frac{\Lambda(i\alpha)}{\Lambda(- i \alpha)} \, y^{-i\alpha} \right]
  = \frac{\sqrt{y}}{2\pi} \int_{\mathbb{R}} d\alpha  \; z_{\frac{1}{2}+i\alpha}\, y^{i\alpha} \,,
 \label{eq:Z0fluctRepRep}
\end{equation}
where we used $\Lambda(s)E_s(\tau) = \Lambda(1-s)E_{1-s}(\tau)$. The spin 0 partition function is thus a Mellin transform of $z_{\frac{1}{2}+i\alpha}$. We now assume the ramp in the spin $0$ sector: 
\begin{equation}
\big\langle \widetilde{Z}_\text{P,cont.}^0(y_1) \widetilde{Z}_\text{P,cont.}^0(y_2) \big\rangle_\text{ramp}  = \frac{1}{2\pi} \, \frac{y_1 y_2}{y_1+y_2} + \ldots \qquad (y_i \gg 1, \; \frac{y_1}{y_2} = \text{fix})\,.
\label{eq:Z0Pramp}
\end{equation}
We can use this spin 0 expression to compute its contribution to the overlap with the Eisenstein series using the unfolding trick \eqref{eq:overlap-coeff}:
\begin{equation}
\begin{split}
    \big\langle z_{\frac{1}{2}+i\alpha_1}\, z_{\frac{1}{2}+i\alpha_2} \big\rangle
   &= \int_0^\infty dy_1'  dy_2' \; (y_1')^{-\frac{3}{2}-i\alpha_1} (y_2')^{-\frac{3}{2}-i\alpha_2} \, \big\langle \widetilde{Z}_\text{P,cont.}^0(y_1') \widetilde{Z}_\text{P,cont.}^0(y_2') \big\rangle \\
   & \sim \frac{\pi}{\cosh(\pi \alpha_1)} \,\delta(\alpha_1 + \alpha_2)  \qquad (|\alpha_i| \rightarrow \infty)\,,
 \end{split} \label{eq:alpha-corr-spin-0}
\end{equation}
where we used the ansatz \eqref{eq:Z0Pramp}. The result \eqref{eq:alpha-corr-spin-0} is obtained by performing an exact Mellin transform of \eqref{eq:Z0Pramp}, disregarding any subleading corrections.
The fact that this is really only an asymptotic condition for large $|\alpha_i|$ has two aspects to it:\footnote{ We thank E.\ Perlmutter for pointing out subtleties with the analytic continuation. See also \cite{DiUbaldo:2023qli} for a detailed discussion.} transforming \eqref{eq:alpha-corr-spin-0} back to $y_i$-variables, we evaluate the constraint $\delta(\alpha_1+\alpha_2)$, obtaining a single integral:
\begin{equation}
   \big\langle \widetilde{Z}_\text{P,cont.}^0(y_1) \widetilde{Z}_\text{P,cont.}^0(y_2) \big\rangle_\text{ramp}
   \approx \frac{\sqrt{y_1y_2}}{4\pi} \int_\mathbb{R} d\alpha \, e^{i\alpha  \log(y_1/y_2)}  \, \frac{1}{\cosh(\pi \alpha)} \,.
\end{equation}
While this can be evaluated explicitly (giving \eqref{eq:Z0Pramp}), let us understand the dominant contribution to the integral.
In the Euclidean limit $y_i\rightarrow \infty$ with $\frac{y_1}{y_2} = \text{fix}$, the integral is dominated by small $\alpha$ and a saddle point at $\alpha=0$ reproduces \eqref{eq:alpha-corr-spin-0} up to subleading corrections. However, the Lorentzian spectral form factor implements the analytic continuation $y_{1,2} \rightarrow \beta \pm i T$ with $T\gg \beta \gg 1$. In this limit, we find $\log(y_1/y_2) \rightarrow \pm \pi i$. The integral then receives contributions from $\alpha \rightarrow \mp \infty$ and the large $|\alpha|$ asymptotics of \eqref{eq:alpha-corr-spin-0} is the crucial ingredient needed to reproduce the linear ramp. 

\paragraph{Spin $m>0$ ramp in $\xi$ variables:} We now analyze the spin $m$ ramp. Using \eqref{eq:Zmfluct}, we have
\begin{equation}
\label{eq:SFFdefyxi}
    \big\langle \widetilde{Z}_\text{P,cont.}^m(y_1)\widetilde{Z}_\text{P,cont.}^m(y_2) \big\rangle = \frac{\sqrt{y_1y_2}}{4\pi^2} \int d\xi_1 d\xi_2 \, e^{-2\pi m(y_1 \cosh \xi_1 + y_2 \cosh \xi_2)} \,\big\langle {\cal Z}^m_\text{P,cont.}(\xi_1) {\cal Z}^m_\text{P,cont.}(\xi_2) \big\rangle
\end{equation}
Assuming the spin $m$ ramp and taking $y_i\rightarrow \infty$, $\frac{y_1}{y_2}=\text{fixed}$ thus corresponds to taking the saddle-point $\xi_{1,2} \rightarrow 0$ ($\xi_1 \approx \xi_2$). Thus, the ramp in the usual $y_i$ variables corresponds to a specific kind of correlations at small $\xi_i$, namely:
\begin{equation}
\label{eq:xiRamp}
    \big\langle {\cal Z}^m_\text{P,cont.}(\xi_1) {\cal Z}^m_\text{P,cont.}(\xi_2) \big\rangle \approx - \frac{\xi_1^2+\xi_2^2}{(\xi_1^2-\xi_2^2)^2} + \ldots \qquad (\xi_i \rightarrow 0) \,,
\end{equation}
which holds for any spin $m$. This is easily verified by plugging into \eqref{eq:SFFdefyxi} and evaluating by saddle point for large $y_i$. This looks similar to eigenvalue repulsion in traditional energy variables, particularly when we take $|\xi_1-\xi_2|\ll\xi_i$, in which case \eqref{eq:xiRamp} becomes $-\frac{1}{2(\xi_1-\xi_2)^2}$; this is actually expected, as the spectral form factor is related to correlations in the density of states as
\begin{equation}
    \big\langle \widetilde{Z}^m_{\text{P}}(y_1) \widetilde{Z}^m_{\text{P}}(y_2) \big\rangle=\frac{\sqrt{y_1 y_2}}{e^{\frac{\pi}{6}(y_1+y_2)}}\int_{E_m}^{\infty} dE_1 dE_2 \, \langle \widetilde{\rho}_\text{P}^{\,m}(E_1)\widetilde{\rho}_\text{P}^{\,m}(E_2)\rangle \,e^{-\left(y_1 E_1+y_2 E_2 \right)} \,,
\end{equation}
where $\widetilde{\rho}_\text{P}^{\,m}(E)$ is the  density of spin $m$ primary states counted by $\widetilde{Z}_\text{P}^m$.\footnote{Thus, when neglecting the Maass cusp forms, the $\xi$ variables are related to energy via a simple (but spin-dependent) change of variables:
\begin{equation}
    {\cal Z}^m_\text{P}(\xi) = 2\pi^2 m |\sinh\xi| \; \widetilde{\rho}_\text{P}^{\,m} \left(2\pi m\cosh\xi-\frac{\pi}{6}\right)\,.
\end{equation}
Taking the appropriate limits, traditional eigenvalue repulsion in energy variables takes the same form in $\xi$ variables.}

To summarize, the universal ramp, which describes the regime of small energy differences, equivalently describes the asymptotics of correlations in $z_{\frac{1}{2}+i\alpha}$ and the small $\xi$ statistics of ${\cal Z}^m_\text{P,cont.}(\xi)$. 

\paragraph{Independence of the ramps:}

We now use \eqref{eq:Zmpdisc} and \eqref{eq:K-kernel} to see how this information is related across different spin sectors. We focus on the imprint of spin $m$ data onto the spin $1$ statistics, showing that each ramp relies on different regimes of $\xi_i$ in the  spin 1 correlations. 

First, we numerically analyze the imprint of the spin 0 ramp on the spin $m>0$ spin sectors in section \ref{sec:numerics}, and find that it gives a subleading correction. We conclude that the spin 0 ramp is independent of the ramp in other spin sectors.

We now look at how the spin $m>0$ ramps are related to each other by focusing on how they appear in the spin 1 correlations.
For transformations between spin $m$ and spin 1, the kernels \eqref{eq:calZcalZtrf} transforming between spin sectors take a simple form:
\begin{equation}
\label{eq:kernelSpin1}
    \begin{split}
    {\cal K}_\text{cont.}^{m,1}(\xi,\xi') &= \frac{1}{4}  \sum_\pm \sum_{d|m}\delta \left(  \xi \pm \xi' \pm \log \frac{d^2}{m} \right) \,,
    \end{split}
\end{equation}
This yields the spin $m$ correlations in terms of spin 1 correlations:
\begin{equation}
\label{eq:spin-m-1-statistics}
\boxed{
    \left\langle{\cal Z}^m_\text{P,cont.}(\xi_1){\cal Z}^m_\text{P,cont.}(\xi_2)\right\rangle = \frac{1}{16}\sum_\pm \sum_{d_1,d_2|m} \left\langle {\cal Z}^1_\text{P,cont.}\!\left[ \pm \xi_1 \pm \log \frac{d_1^2}{m} \right]{\cal Z}^1_\text{P,cont.}\!\left[ \pm \xi_2 \pm \log \frac{d_2^2}{m} \right]\right\rangle
}
\end{equation}
This is an exact statement. It implies how a ramp of the l.h.s.\ would be encoded in the spin 1 statistics: it corresponds to a particular behavior for small $\xi_i$, which is equivalent to a sum of spin 1 correlations around $\xi = \pm \log\left(\frac{d^2}{m}\right)$ for all $d|m$. 

This immediately proves what we set out to show: for example, the existence of a spin 1 ramp is insufficient to conclude the existence of the ramp for any other spin. For any $m>1$, there is some divisor such that $\log\frac{d^2}{m}$ is outside the small $\xi$ regime that the spin 1 ramp provides information for. In general, the existence of a spin $m'$ ramp cannot be used to find the ramp for any other spin, as any ramp only gives a particular linear combination of spin 1 correlations. 

Further, any finite collection of spins, with spin $m$ being the largest, can only give information on correlations of ${\cal Z}^1_\text{P,cont.}[\xi]$ for $|\xi|<\log m$. This is insufficient to find the ramp for any spin $m'>m$, which requires knowledge of correlations for $\log m < |\xi_{1,2}|<\log m'$.\footnote{For example, assume a ramp for every spin up to $m=10$; this only gives us information about the spin 1 correlations for a subset of $\xi_{1,2}\in [-\log10,\log10]$; thus, we wouldn't have all the necessary information to find the ramp for any spin $m'>10$.} Therefore, assuming universal random matrix statistics in any finite collection of spin sectors is insufficient to conclude the same for all other spins just based on symmetries.

\vspace{5pt}

\vspace{5pt}
\subsection{Signatures of the spin \texorpdfstring{$m$}{m} ramp in other spin sectors (numerically)}\label{sec:numerics}

Armed with the abstract argument of the previous subsection, we now put this on a firmer footing by numerically analyzing the imprint of the spin $0$ and spin $1$ ramps on the other spin sectors. This analysis will involve extrapolating the functional form of the ramp outside of the regime where it is strictly valid (large $y$, or small $\xi$). It should thus be taken with a grain of salt; however, our results will be consistent with the abstract argument given above, which we take as evidence that the analysis is justified {\it a posteriori}.

\vspace{5pt}
\subsubsection{Signatures of a spin \texorpdfstring{$m=0$}{m=0} ramp}
\label{sec:spin0imprint}

We begin with the imprints of universal correlations in the near-extremal spin 0 spectrum onto other spin sectors and ask the following question: can we infer a ramp in the near-extremal spin $m$ spectrum, by just assuming the existence of a ramp in the near-extremal spin 0 spectrum (subject to the caveat mentioned above)? We will give evidence that the answer is `no'.

We begin with the `ramp' in the spin 0 sector, which was given in \eqref{eq:Z0Pramp}. Such a spin 0 ramp determines the correlations of the overlap coefficients with Eisenstein series for large $\alpha$; this was given in  \eqref{eq:alpha-corr-spin-0}, which we reproduce here fore convenience:
\begin{equation}
\begin{split}
    \big\langle z_{\frac{1}{2}+i\alpha_1}\, z_{\frac{1}{2}+i\alpha_2} \big\rangle
   & \supset_{_0} \int_0^\infty dy_1'  dy_2' \; (y_1')^{-\frac{3}{2}-i\alpha_1} (y_2')^{-\frac{3}{2}-i\alpha_2} \, \big\langle \widetilde{Z}_\text{P,cont.}^0(y_1') \widetilde{Z}_\text{P,cont.}^0(y_2') \big\rangle \\
   & = \frac{\pi}{\cosh(\pi \alpha_1)} \,\delta(\alpha_1 + \alpha_2) + \ldots 
 \end{split}
\end{equation}
where ``$\supset_{_0}$'' means that we only consider the contribution to the l.h.s., which comes from the ramp at spin 0, \eqref{eq:Z0Pramp}. This, of course, only contains partial information about the true microscopic spectrum.
We wish to confirm that this partial information is insufficient to deduce the existence of a universal ramp at spin $m$.
To this end, we use the result to compute the Eisenstein series contribution to the spectral form factor in the spin $(m_1,m_2)$ sector:
\begin{equation}
\label{eq:Zm1Zm2new}
\begin{split}
    &\big\langle \widetilde{Z}^{m_1}_\text{P,cont.}(y_1) \widetilde{Z}^{m_2}_\text{P,cont.}(y_2)
    \big\rangle = \frac{1}{(4\pi)^2}  \int d\alpha_1 d\alpha_2 \, \big\langle z_{\frac{1}{2}+i\alpha_1}\, z_{\frac{1}{2}+i\alpha_2} \big\rangle \,E_{\frac{1}{2} + i \alpha_1}^{m_1}(y_1)\, E_{\frac{1}{2} + i \alpha_2}^{m_2}(y_2)
 \end{split}
 \end{equation}
We analyze separately the cases where both spins are non-zero, and where one of them is zero.
\paragraph{(a) Both spins $m_1, m_2\geq 1$:} When neither spin in \eqref{eq:Zm1Zm2new} is zero, we have
\begin{equation}
\begin{split}
    &\big\langle \widetilde{Z}^{m_1}_\text{P,cont.}(y_1) \widetilde{Z}^{m_2}_\text{P,cont.}(y_2)
    \big\rangle 
    \\
    &\quad \supset_{_0}
     \frac{\sqrt{y_1y_2} }{\pi} 
    \int d\alpha \, \frac{\sigma_{2i\alpha}(m_1)\sigma_{-2i\alpha}(m_2)}{m_1^{i\alpha}m_2^{-i\alpha}\Lambda(i\alpha)\Lambda(-i\alpha)} \frac{K_{i\alpha}(2\pi m_1 y_1)K_{-i\alpha}(2\pi m_2 y_2)}{\cosh(\pi\alpha)} 
\\    &\quad  = \frac{\sqrt{y_1y_2} }{\pi^2} 
   \sum_{\substack{d_1|m_1\\d_2|m_2}} \int d\alpha \, \frac{\alpha \tanh(\pi\alpha)}{|\zeta(2i\alpha)|^2}  \left( \frac{m_2d_1^2}{m_1d_2^2} \right)^{i\alpha} K_{i\alpha}(2\pi m_1 y_1)K_{-i\alpha}(2\pi m_2 y_2)
\end{split}
\label{eq:ZmZmInt}
\end{equation}
The integrand has poles at the following locations:
\begin{equation}
    \text{poles:} \qquad\quad \alpha = \pm \frac{i}{2} \, \eta_n \equiv \pm \frac{i}{4} \mp \frac{1}{2} \, \text{Im}(\eta_n) \,,\qquad\qquad \alpha = \frac{(2k-1)i}{2} \;,\; k\in \mathbb{Z}
\end{equation}
where $\eta_n = \frac{1}{2} \pm i \{ 14.135\cdots ,\, 21.022\cdots ,\, 25.011\cdots , \ldots \}$ are the non-trivial zeros of the $\zeta$-function. The second set of zeros is due to the factor $1/\cosh(\pi \alpha)$. The integrand falls off exponentially in all directions, so we can evaluate the integral by closing the contour either at $+i\infty$ or $-i\infty$. By Cauchy's theorem, we get
\begin{equation}
\begin{split}
    \big\langle \widetilde{Z}^m_\text{P,cont.}(y_1) \widetilde{Z}^m_\text{P,cont.}(y_2)
    \big\rangle 
    &\supset_{_0}
    2\pi i \left[\, \sum_{\substack{\eta_n\\ \text{Im}(\eta_n)>0}} \text{Res}_{\alpha= \frac{i}{2} \eta_n} \left( \cdots \right) + \sum_{k\geq 1} \text{Res}_{\alpha=\frac{(2k-1)i}{2}} \left( \cdots \right) \right]
\end{split}
\end{equation}
where $(\cdots)$ is the integrand in \eqref{eq:ZmZmInt}, including all the pre-factors. One can easily see that the values of these residues decay quickly as $|\eta_n|$ (for the first sum) or $k$ (for the second sum) increases. It is therefore straightforward to evaluate these sums numerically by including sufficiently many terms.\footnote{ Roughly, including ${\cal O}(10\!-\!100)$ terms is sufficient to get an accuracy of ${\cal O}(10^{-5})$.}

We find (numerically) for the contribution to the spin $(m_1,m_2)$ spectral form factor due to the existence of a ramp at spin 0:
\begin{equation}
\label{eq:ZZresNum}
\boxed{
\big\langle \widetilde{Z}^{m_1}_\text{P,cont.}(y_1) \widetilde{Z}^{m_2}_\text{P,cont.}(y_2)
    \big\rangle \supset_{_0} \lambda_{m_1} \, \delta_{m_1 m_2} \; e^{-2\pi (m_1 y_1+m_2y_2)} \, \sqrt{\frac{y_1 y_2}{y_1+y_2}}  + \ldots 
    \quad\;\; (y_i \gg 1)
    }
\end{equation}
where the first few spin-dependent prefactors are 
\begin{equation}
    \begin{split}
        \lambda_1 = 0.761.. \,,\quad
        \lambda_2 = 0.644.. \,,\quad
        \lambda_3 = 0.613.. \,,\quad
        \lambda_4 = 0.532.. \,,\quad
        \lambda_5 = 0.548.. \,,\;\; \text{etc.} 
    \end{split}
\end{equation}
Note that these proportionality constants are non-monotonic in spin due to their complicated dependence on number-theoretic properties of the spin. We show some of the numerical analysis leading to this conclusion in appendix \ref{app:numerics}.
Note that, given the simple universal form of \eqref{eq:ZZresNum}, it is tempting to speculate that there might be an analytical argument to prove it.

Most importantly, \eqref{eq:ZZresNum} is {\it not} the form expected for a ramp (compare the non-exponential piece to \eqref{eq:Z0Pramp}). It should instead be thought of as a subleading (in large $y_i$) correction to the spin $m$ ramp. This shows that the imprint of the ramp in the spin 0 sector onto a higher spin sector is not a ramp, but merely a subleading correction to it. {\it Any potential ramp in the higher spin sector is therefore not due to a ramp at spin 0 plus symmetries.} This strengthens the analytic argument of the previous section.

\paragraph{(b) One spin $m_1 > 0$ and one vanishing spin $m_2=0$.} 
Finally, if one spin is 0, we have
\begin{equation}
\begin{split}
    \big\langle \widetilde{Z}^{m_1}_\text{P,cont.}(y_1) \widetilde{Z}^{0}_\text{P,cont.}(y_2)
    \big\rangle & \supset_{_0}
     \frac{\sqrt{y_1y_2} }{2\pi} 
    \int d\alpha \, \frac{\sigma_{2i\alpha}(m_1)}{m_1^{i\alpha}\Lambda(-i\alpha)} \frac{K_{i\alpha}(2\pi m_1 y_1) y_2^{-i\alpha}}{\cosh(\pi\alpha)} 
\\    &  = \frac{\sqrt{y_1y_2} }{2\pi} 
   \sum_{\substack{d_1|m_1}} \int d\alpha \, \frac{ K_{i\alpha}(2\pi m_1 y_1)y_2^{-i\alpha}}{\Gamma(-i\alpha)\zeta(-2i\alpha) \cosh(\pi \alpha)} \left( \frac{d_1^2}{m_1\pi}\right)^{i\alpha}
\end{split}
\label{eq:ZmZ0Int}
\end{equation}
This can be analyzed in the same way as \eqref{eq:ZmZmInt} by residues. Unsurprisingly, one finds very small values (compared to those expected for a ramp), similar to the other unequal spin cases shown in figures \ref{fig:ZmZm1} and \ref{fig:ZmZm2}.

\vspace{5pt}
\subsubsection{Signatures of a spin \texorpdfstring{$m=1$}{m=1} ramp}
\label{sec:spin1imprint}

We have argued in section \ref{sec:ramp-independence} that for every $m$, ${\cal Z}^m_\text{cont.}(\xi)$ contains information about ${\cal Z}^1_\text{P,cont.}(\xi)$ that is not contained in any ${\cal Z}^{m'}_\text{P,cont.}(\xi)$ with $m'<m$. For illustration, we shall now make this abstract argument more concrete by repeating the analysis of section \ref{sec:spin0imprint} for higher spin (which involves different manipulations): we shall assume the existence of universal eigenvalue repulsion in the spin $m=1$ sector and analyze its imprint on other spin $(m_1,m_2)$ sectors via spectral determinacy arguments (i.e., symmetries).

The universal ramp ansatz in the spin $m$ sector is given by \eqref{eq:xiRamp} in the $\xi$ variables.
The imprint of that expression onto the spin $(m_1,m_2)$ sector is:
 \begin{equation}
  \begin{split}
      &\big\langle \widetilde{Z}_\text{P,cont.}^{m_1}(y_1)\widetilde{Z}_\text{P,cont.}^{m_2}(y_2) \big\rangle \\ 
     &\qquad\supset_{_m} \frac{\sqrt{y_1y_2}}{4\pi^2} \iint d\xi_1 d\xi_2 \, e^{-2\pi(m_1y_1\cosh \xi_1 + m_2y_2 \cosh \xi_2)} \\
     &\qquad\qquad\qquad\qquad \times \iint d\xi_1' d\xi_2' \, {\cal K}^{m_1,m}_\text{cont.}(\xi_1,\xi_1') {\cal K}^{m_2,m}_\text{cont.}(\xi_2,\xi_2') \,\left[ -\frac{\xi_1'^2+\xi_2'^2}{(\xi_1'^2 - \xi_2'^2)^2}\right]
  \end{split}
 \end{equation}
 where the symbol ``$\supset_{_m}$'' means that we only consider the contribution to the l.h.s.\ originating from a ramp at spin $m$.\footnote{Again, we extrapolate the validity of the ramp (valid at small $\xi$), which is only justified by the consistent result we find.}
 For simplicity consider $m=1$, where the kernels ${\cal K}^{m_i,m}_\text{cont.}$ reduce to a sum over delta-functions:
  {\small
  \begin{equation}
  \begin{split}
      &\big\langle \widetilde{Z}_\text{P,cont.}^{m_1}(y_1)\widetilde{Z}_\text{P,cont.}^{m_2}(y_2) \big\rangle \\ 
     &\quad 
     \supset_{_{m=1}} -\frac{\sqrt{y_1y_2}}{16\pi^2} \sum_{{\substack{ d_1|m_1 \\ d_2|m_2}}} \sum_\pm \iint d\xi_1 d\xi_2 \,  \frac{\left(\xi_1  \pm \log(d_1^2/m_1) \right)^2 + \left(\xi_2  \pm \log(d_2^2/m_2) \right)^2 }{\left( \left(\xi_1  \pm \log(d_1^2/m_1) \right)^2 - \left(\xi_2  \pm \log(d_2^2/m_2) \right)^2\right)^2} \, e^{-2\pi (m_1y_1\cosh \xi_1 + m_2y_2 \cosh \xi_2)}
  \end{split}
  \label{eq:ZmZmspin1Imprint}
 \end{equation}
 }\normalsize
 For large $y_i$ the integrals are dominated by a saddle point at small $\xi_i$. This can be straightforwardly evaluated numerically, and we find (within numerical accuracy):
  \begin{equation}
  \label{eq:Zm1Zm1result}
  \boxed{
      \big\langle \widetilde{Z}_\text{P,cont.}^{m_1}(y_1)\widetilde{Z}_\text{P,cont.}^{m_2}(y_2) \big\rangle
     \supset_{_{m=1}}  \sigma_0(m_1) \times\delta_{m_1m_2} \, \frac{1}{2\pi}  \,\frac{y_1y_2}{y_1+y_2} \,  e^{-2\pi m_1 ( y_1 +  y_2)}
  }
 \end{equation}
where $\sigma_0(m)$ is the number of divisors of $m$ (including $1$ and $m$). The source of this prefactor are the terms with $d_1=d_2$. Some of our numerical analysis is again shown in appendix \ref{app:numerics} (in particular figure \ref{fig:ZmZm3}). That is, the imprint of a ramp at spin 1 onto the spin $m$ sector is $\sigma_0(m)$ times the ramp.\footnote{We have not analyzed the imprint of the spin $m>0$ ramps on the spin 0 sector.} For our conclusions to hold, the factor $\sigma_0(m)$ is crucial:\footnote{ As in the spin 0 case the numerical agreement is excellent, so one might expect there to be an analytical way to evaluate \eqref{eq:ZmZmspin1Imprint}.} the functional dependence on $y_i$ is indeed that of a ramp, so the discrepancy with random matrix universality is only due to the mismatching prefactor. However, this prefactor is always at least 2, so it never matches the random matrix theory expectation (which involves not just a linear ramp, but also a very specific normalization thereof). This is sufficient to conclude that a ramp in the spin 1 Eisenstein sector does not imply random matrix universality in higher spin sectors.

\vspace{10pt}
\section{Cardy-like constraints on spectral correlations}
\label{sec:cardy}

With the knowledge that the ramp in each spin sector does indeed contain independent data despite spectral determinacy, we now take this as a starting point to find more information about the spectrum. Following the derivation of the Cardy formula, we will find that the existence of eigenvalue repulsion in the near-extremal spectrum predicts the same for a part of the spectrum far from extremality.

\vspace{5pt}
\subsection{Review: modular invariance of the density of states}
While modular invariance is most naturally formulated as an invariance of the partition function, following, e.g., \cite{Maxfield:2019hdt, Zamolodchikov:2001ah} we can write it as an invariance of the density of states of primaries. First, we define variables $(P,\bP)$ that are more suitable for applying modular transformations to the density of states directly:
 \begin{equation}
     h=\frac{c-1}{24}+P^2\,,\quad \bh=\frac{c-1}{24}+\bP^2\,. \label{eq:Pvars}
 \end{equation}
 The energy and spin variables we used previously are
 \begin{equation}
 E \equiv 2\pi\left(h+\bar{h}-\frac{c}{12}\right) \equiv 2\pi \left(P^2 + \bP^2 -\frac{1}{12} \right) \,,\qquad \mc \equiv h-\bar{h} \equiv P^2 - \bar{P}^2\,. \label{eq:energy-p-varchange}
 \end{equation}
It will be convenient to allow for positive and negative spins in this section. In addition, we define $\mc \in \mathbb{R}$ as the continuous spin, via $\rho(E)=\sum_{m\in \mathbb{Z}} \,\rho^m(E) =\int d\mc\, \rho(E,\mc)$.
The $P$ variables can take both imaginary and real values, corresponding to different parts of the spectrum:
\begin{equation}
\begin{split}
  \text{censored states:} \qquad &0< -iP \leq \sqrt{\frac{c-1}{24}}  \qquad \Leftrightarrow \qquad 0\leq h \leq \frac{c-1}{24} 
  \\
  \text{dense states:} \qquad & 0 < P \qquad\qquad\qquad \quad\,\;\;\Leftrightarrow  \qquad h > \frac{c-1}{24}  
\end{split}
\end{equation}
and similarly for $\bar{h}$.
With these variables, we can write the CFT partition function as 
\begin{equation}
    Z(\tau,\bar{\tau})=\int_{-\infty}^{\infty}\frac{dP}{2} \frac{d\bP}{2} \, \rho_\text{P}(P,\bP)\chi_\text{P}(\tau) \bar{\chi}_{\bar{\text{P}}}(\bar{\tau}) \,,\qquad \chi_\text{P}(\tau)=\frac{e^{2\pi i \tau P^2}}{\eta(\tau)}\,,\;\;
    \bar{\chi}_{\bar{\text{P}}}(\bar\tau)=\frac{e^{-2\pi i \bar\tau \bP^2}}{\eta(-\bar\tau)}\,,
    \label{eq:Z_in_Pvars}
\end{equation}
where $\rho_\text{P}(P,\bP)$ is the density of states of primary operators (we allow $(P,\bP)$ to be negative, with $\rho_\text{P}(P,\bP)$ being even), and the Virasoro characters account for descendant states. As before, $\rho_\text{P}(P,\bP)$ is a sum of delta-functions, which now includes support on imaginary values of $P$ \cite{Maxfield:2019hdt}. 

In these variables, invariance of $Z$ under the modular S-transform can be recast as the invariance of the density of states under a Fourier transform:
\begin{equation}
\label{eq:Sinvariance}
    \rho_\text{P}(P,\bP) = 2\int_{-\infty}^{\infty} dP' d\bP' \,  e^{-4\pi i (P P'+\bP \bP')}\rho_\text{P}(P',\bP')\,.
\end{equation}
This formulation of modular invariance makes the Cardy formula and the form of the asymptotic density of states in the lightcone limit ($\bP>0 \text{ fixed}, \, P\rightarrow \infty$)  more transparent; rather than having to work with the partition function (and complicated questions of convergence \cite{Mukhametzhanov:2020swe}) we can directly compute with the density of states. 

For illustration, let us recall the derivation of the Cardy formula in these variables \cite{Maxfield:2019hdt}: the contribution of the vacuum state to the density of states is $\rho_\text{P}(P,\bP) \supset \rho_\text{P,vac}(P) \rho_\text{P,vac}(\bP)$ with
\begin{equation}
       \rho_\text{P,vac}(P) = \left[ \delta\left( P - i\,\sqrt{\frac{c-1}{24}} \right)  - \delta\left( P - i \sqrt{ \frac{c-1}{24}-1} \right) \right] \;+\; (P \leftrightarrow -P )\,,
\end{equation}
where the subtraction accounts for the null descendants of the vacuum. The modular S-transform of this contribution according to \eqref{eq:Sinvariance} gives the leading contribution to the density of states as $P,\bP\rightarrow \infty$, i.e., the Cardy formula:
\begin{equation}
     \rho_\text{P}(P,\bP) \approx  2 \exp\left(2\pi (P+\bP)\sqrt{\frac{c-1}{6}}\right) \qquad\quad(P,\bP\rightarrow \infty)\,.
\end{equation}

\vspace{5pt}
\subsection{S-dual of eigenvalue repulsion}\label{sec:sdual-ramp}

Using the same ansatz for the primary partition function as section \ref{sec:ramp}, we assume the near-extremal density of primaries exhibits random matrix statistics. The connected two-point function of the density of states is
\begin{equation}
    \langle \rho_\text{P}(P_1',\bP_1')\rho_\text{P}(P_2',\bP_2')\rangle =  \sum_{m_1,m_2\in \mathbb{Z}}  F^{m_1,m_2}(P_1',\bP_1',P_2',\bP_2') \,
  \delta(P_1'^2-\bP_1'^2-m_1)\delta(P_2'^2-\bP_2'^2-m_2).
     \label{eq:rho-2pt-pavars}
\end{equation}
In the near-extremal limit, we obtain the ramp by applying the change of variables in \eqref{eq:energy-p-varchange} to the expression for universal eigenvalue repulsion in \eqref{eq:ramp-energy}:\footnote{ This expression involves the Jacobian $\big| \frac{\partial(E_i,\mc_i)}{\partial(P_i',\bP_i')}\big| =16\pi|P_i' \bP_i'|$.}
\begin{equation}
    F^{m_1,m_2}(P_1',\bP_1',P_2',\bP_2')\approx -\frac{1}{2\pi^2}\frac{64|P_1'\bP_1' P_2' \bP_2'|}{\left(P_1'^2 + \bP_1'^2-P_2'^2-\bP_2'^2\right)^2}\,\delta_{m_1,m_2} \qquad\; \left( \begin{aligned} & P_i',\bP_i' \text{ near-extremal}\\& \quad\;\omega\ll E_i -E_{m_i}
    \end{aligned}\right)\label{eq:ramp-p-vars}
\end{equation}
which is expected to hold near extremality, defined by the
\begin{equation}
\label{eq:extremalP}
\text{near-extremal regime:} \qquad\quad
   0<\text{min}\left(P_i^2,\bP_i^2\right) \ll 1 
\end{equation}
and we also take $\omega \equiv2\pi( P_1'^2 + \bP_1'^2-(P_2'^2+\bP_2'^2)) \ll E_{i} -E_{m_{i}}=4\pi \, \text{min}\left(P_i^2,\bP_i^2\right)$,
where we are using the expression for the small $\omega$ limit as in \eqref{eq:extremalP}.\footnote{We also assume that the divergence as $\omega \rightarrow 0$ of the gravity and RMT correlations is regulated (in RMT, the sine kernel regulates the divergence; in gravity, presumably more complicated wormhole configurations regulate the divergence).\label{foot:reg}}

We now take the modular S-transform of \eqref{eq:ramp-p-vars}, first with respect to $P_1', \, \bP_1'$, then with respect to the second set of variables. The S-transform is a Fourier transform, which we will evaluate by saddle point. Because we want the saddle point $(P_{1*}',\bP_{1*}')$ to lie in the near-extremal regime where \eqref{eq:ramp-p-vars} applies, the S-dual parameters $(P_i, \, \bP_i)$ should lie {\it outside} the near-extremal limit. We achieve this by identifying the
\begin{equation}
\label{eq:farExtremal}
 \quad\text{large spin, far-from-extremality regime:} \qquad\quad   P_i^2\gg\bP_i^2\gg 1 \qquad (\,\Leftrightarrow \; \mc_i\gg \tau_i \gg 1\,)\quad
\end{equation}
where, here and in the following, we take w.l.o.g.\ $\mc_i > 0$, and $P_i^2$ parametrically larger than $\bP_i^2$.\footnote{I.e., $P^2=\Lambda P_0^2, \, \bP_2=\Lambda^v \bP_0^2$, with $\Lambda \rightarrow \infty, \, 0<v<1$.} (In appendix \ref{sec:saddle-details} we give analogous expressions valid for any choice of signs for $\mc_i$.)
In the bracket, we have written the condition in terms of the spin and (shifted) twist $\tau_i\equiv \frac{E_i}{2\pi}-\mc_i=2\bP_i^2-\frac{1}{12} \approx 2\bP_i^2$, thus showing how this regime translates into  energy-spin variables. This regime is one of the most natural to consider when utilizing the S-transform \cite{Pal:2019zzr}; it appears when we allow the conformal dimensions $h, \, \bar{h}$, or equivalently spin and twist, to approach infinity at different rates. Note that this is a distinct regime from that considered by the lightcone bootstrap \cite{Kusuki:2018wpa}, where the twist is bounded by extremality, $\tau\leq -\frac{1}{12}$.

\begin{figure}
    \centering
\includegraphics[width=.325\textwidth]{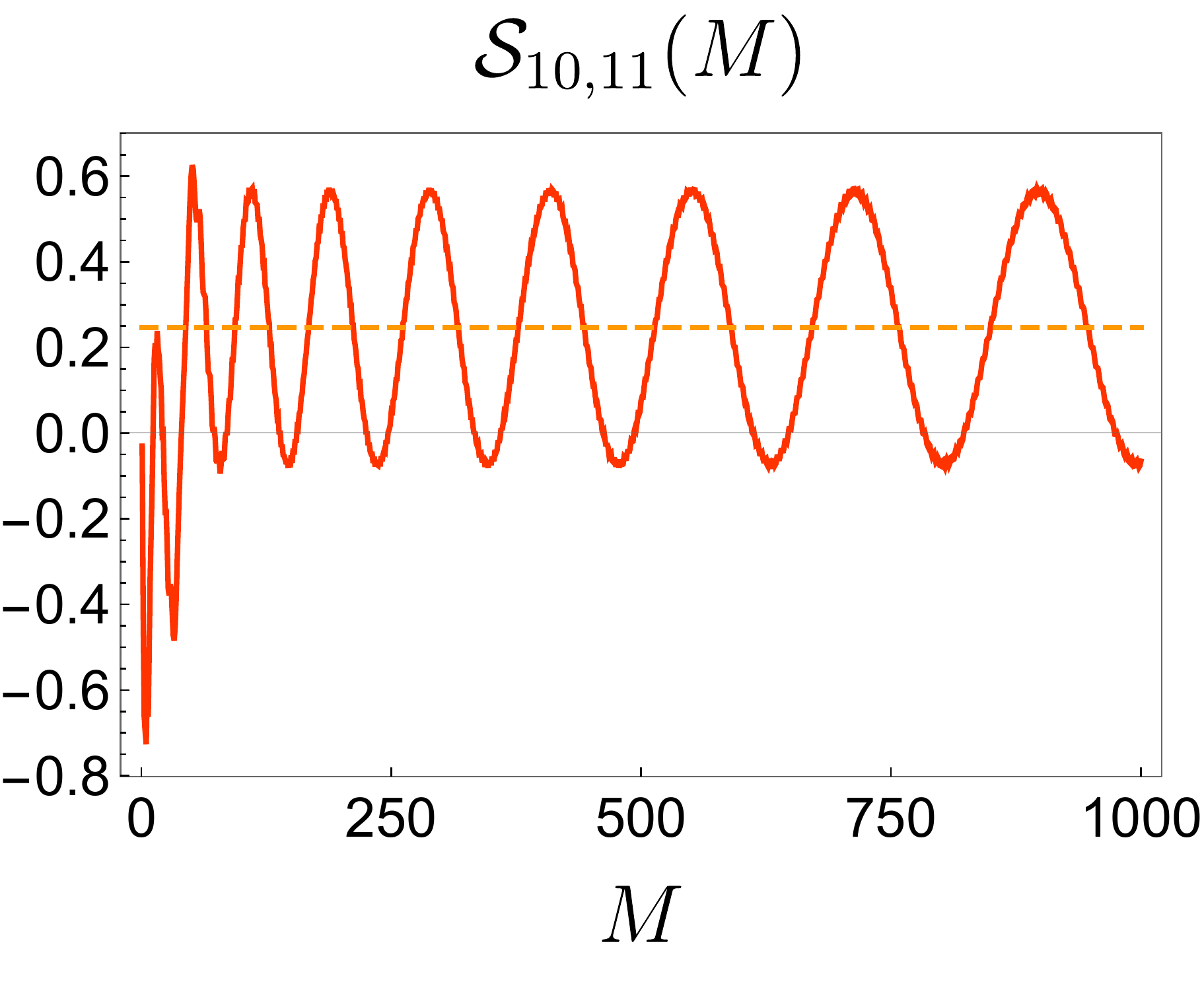}
\includegraphics[width=.325\textwidth]{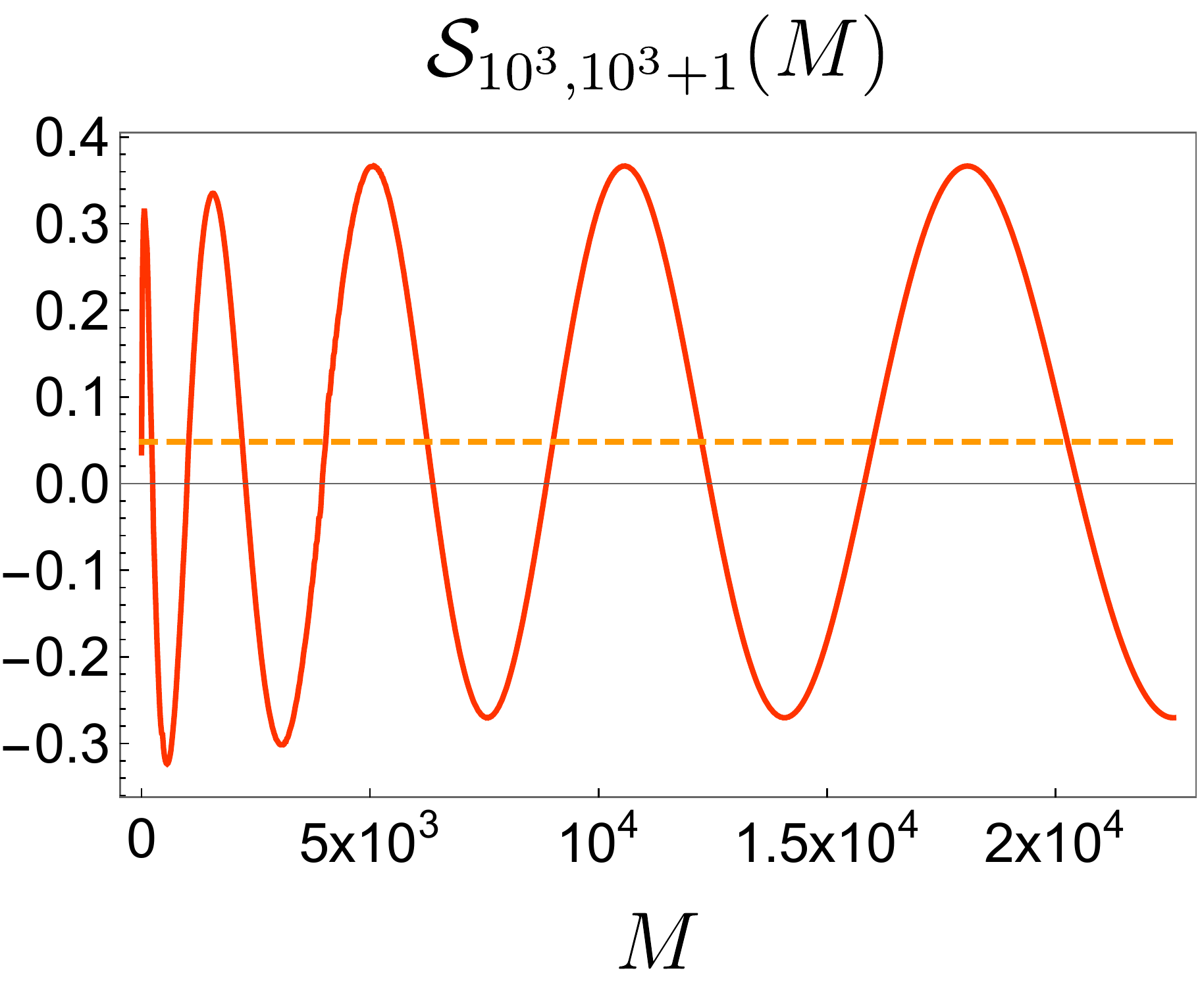}
\includegraphics[width=.325\textwidth]{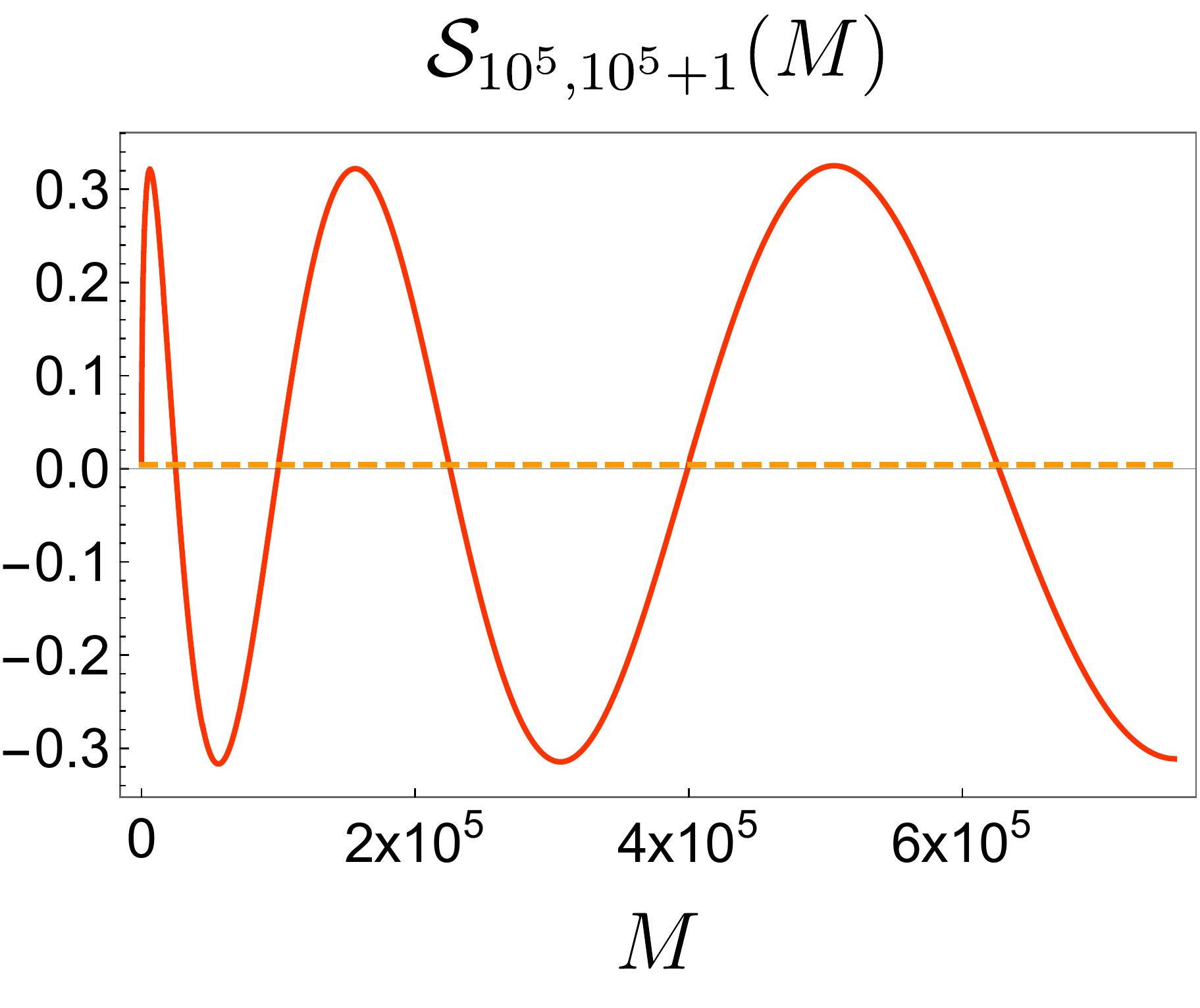}
    \caption{Plots of the non-convergent sum $\mathcal{S}_{\mc_1,\mc_2}(M)$ for $\mc_1 = \mc_2 - 1 \in \{10,10^3,10^5\}$ as a function of the cutoff $M$. We see that, although the limit $M\rightarrow \infty$ does not exist, the function oscillates within a bounded bandwidth. This is in stark contrast to the case $\mc_1 = \mc_2 \equiv \mc $, where ${\cal S}_{\mc,\mc}(M)$ diverges. Note that for large $\mc_i$, the mean value around which the oscillations occur, tends to zero (dashed lines).
    }
    \label{fig:oscillations}
\end{figure}

This choice allows us to evaluate the Fourier transform (S-transform) despite our limited information about the density of states. Concretely, we wish to compute
\begin{equation}
\label{eq:saddleCalc0}
\begin{split}
&\langle \rho_\text{P}(P_1,\bar{P}_1)\rho_\text{P}(P_2,\bar{P}_2)\rangle 
\approx  4\int dP_1' d\bar{P}_1'dP_2' d\bar{P}_2'\;  e^{-4\pi i (P_1' P_1+\bar{P}_1'\bar{P}_1+P_2' P_2+\bar{P}_2'\bar{P}_2)}\\
&\qquad\qquad\qquad\qquad\qquad\quad \times \sum_{m_i\in \mathbb{Z}} F^{m_1,m_2}(P_1',\bar{P}_1',P_2',\bar{P}_2') \,\delta(P_1'^2-\bar{P}_1'^2-m_1)\delta(P_2'^2-\bar{P}_2'^2-m_2)\,.
\end{split}
\end{equation}
For details on the evaluation of these integrals in the regime \eqref{eq:farExtremal}, we refer to appendix \ref{sec:saddle-details}. The result is of the following form:
\begin{equation}
\begin{split}
     &\langle \rho_\text{P}(E_1,\mc_1) \rho_\text{P}(E_2,\mc_2)\rangle \\
     &\qquad \approx 2 \sum_{m_i>0}  F^{m_1,m_2}\left(P_1\sqrt{\frac{m_1}{\mc_1}},\bP_1\sqrt{\frac{m_1}{\mc_1}},P_2\sqrt{\frac{m_2}{\mc_2}},\bP_2\sqrt{\frac{m_2}{\mc_2}}\right)\,\big(\mc_1 \mc_2 m_1 m_2\big)^{-1/4} \\
     &\qquad\qquad\qquad\qquad\quad \times \cos\left(4\pi\sqrt{m_1 \mc_1}+\frac{\pi}{4}\right)\cos\left(4\pi\sqrt{m_2 \mc_2}+\frac{\pi}{4}\right) \qquad \quad (P_i^2 \gg \bP_i^2 \gg 1)\,.
\end{split}\label{eq:s-dual-saddle-preramp}
\end{equation}
Because $\mc_i= P_i^2 - \bP_i^2\gg 1$, we are guaranteed that $\bP_i\sqrt{\frac{m_i}{\mc_i}}$ is small.\footnote{This is strictly true up to an asymptotic regime of large $m_i$, beyond which we assume the contributions are negligible. See appendix \ref{sec:saddle-details} for details.} Ensuring the small $\omega$ condition where \eqref{eq:ramp-p-vars} applies, requires $\frac{\bP_1^2}{P_1^2}-\frac{\bP_2^2}{P_2^2}\ll\frac{\bP_{1,2}^2}{P_{1,2}^2}$. We can then use the expression for $F^{m_1,m_2}$ near extremality, i.e., \eqref{eq:ramp-p-vars}; we find that the connected two-point function for $\mc_i\gg \tau_i\gg 1$ and $E_1-E_2\ll E_i$ is 
\begin{equation}
\label{eq:ssum}
\begin{split}
    \langle \rho_\text{P}(E_1,\mc_1) \rho_\text{P}(E_2,\mc_2)\rangle &\approx -\frac{1}{2\pi^2}\,\frac{\mc_1 \mc_2}{\left(\mc_2 \left(E_1-2\pi\, \frac{c-1}{12}\right) - \mc_1 \left(E_2-2\pi\, \frac{c-1}{12}\right) \right)^2}   \times \lim_{M\rightarrow \infty} {\cal S}_{\mc_1,\mc_2}(M)
\end{split}
\end{equation}
where 
\begin{equation}
\label{eq:Sm1m2Def}
 {\cal S}_{\mc_1,\mc_2}(M) \equiv \frac{1}{(\mc_1 \mc_2)^{1/4}}\;\sum_{m=1}^M \frac{2}{\sqrt{m}}\, \cos\left(4\pi \sqrt{m \mc_1}+\frac{\pi}{4}\right)\cos\left(4\pi \sqrt{m \mc_2}+\frac{\pi}{4}\right) \,.
\end{equation}
The limit $M\rightarrow \infty$ of the expression ${\cal S}_{\mc_1,\mc_2}(M)$ is in fact not well defined without further regularization: as written, the sum does not converge. However, for $m_1 \neq m_2$ it also does not diverge either, but rather oscillates within a bounded window as a function of the cutoff $M$.  Furthermore, the mean value of the oscillation tends to zero as $\mc_i$ grow large.\footnote{ This is also true if $\mc_1$ is held fixed and only $\mc_2$ grows large.} 
We illustrate this in figure \ref{fig:oscillations}.
 We belive that, upon implementing a proper method of coarse-graining over spins, one could derive asymptotic upper and lower bounds for \eqref{eq:Sm1m2Def} by applying techniques similar to the Tauberian theorems deployed in, e.g., \cite{Mukhametzhanov:2019pzy}. However, since the variance we are considering is not strictly positive, we cannot easily apply such theorems and leave this problem to a future analysis. 
 Finally, note that several approximations have been assumed in the computation presented here. A small correction of each term in ${\cal S}_{\mc_1,\mc_2}(M)$ turns this non-convergent series into a convergent one.\footnote{ In particular, any power of $m^{-1}$ in front of the product of cosines that is strictly greater than $1/2$ would produce a convergent sum. Similarly, methods such as Euler summation straightforwardly regularize the sum and give a definite result.} Since  we assumed an approximate expression for the two-point function of the density of states and the latter is in fact a distribution, it is reasonable to expect for the above expressions to be regularized by subleading effects and by integration against test functions. It would be interesting to investigate this more rigorously.

We can get a well defined result for \eqref{eq:ssum} in the decompactification limit: by reintroducing units to spin, $m=\frac{2\pi}{L} n$ where $L$ is the circumference of the cylinder, we replace $\sum_m \rightarrow \frac{L}{2\pi} \int_0^{\infty}dm$ in the limit $L\rightarrow \infty$. The integral over $m$ can then be performed explicitly, and we find:
\begin{equation}
\boxed{
      \langle \rho_\text{P}(E_1,\mc_1) \rho_\text{P}(E_2,\mc_2)\rangle \approx  -\frac{1}{2\pi^2} \frac{1}{\left(E_1 -E_2 \right)^2} \,\frac{L}{2\pi} \,\delta(\mc_1-\mc_2)
      \quad\;\;\; \left(\begin{aligned}|\mc_i|\gg \tau_i \gg 1, \\ |E_1-E_2|\ll E_i \end{aligned}\right)
      }
\end{equation}
This expression encodes eigenvalue repulsion in each spin sector, both the universal functional form and the correct prefactor as expected from random matrix theory. The regime of validity corresponds to the spectrum far from extremality.\footnote{The absence of Dirac delta functions forcing $\mc_i$ to be discrete can be understood using the same reasoning as the lack of Dirac delta functions in the Cardy formula.}

This shows that modular invariance extends the regime in which the statistics of primary operators follows random matrix universality; random matrix statistics for primary operators in the near-extremal spectrum implies the same statistics for primaries far from extremality.

\vspace{10pt}
\section{Discussion}
\label{sec:discussion}

This paper has two main results: first, while spectral determinacy implies very strong constraints between the spectra of primaries in different spin sectors, this does not preclude independent random matrix universality near extremality in each spin sector. Second, modular invariance can be used to argue for random matrix universality in a certain large spin regime far from extremality.

Our goal in the paper was to identify a set of universal quantities for coarse-grained aspects of two-dimensional CFTs with large central charge. We have found that correlations between the coefficients of the modular invariant decomposition of the partition function provide such information (using a certain parametrization for the near-extremal spectrum). This opens the door to explaining these features along the lines of the Efetov sigma model \cite{Wegner,Efetov:1983xg,Altland:2020ccq}. Indeed, in the quantum mechanical case one can write a path integral for the spectral correlators of interest, and in the limit where it is universal this path integral is dominated by an ergodic mode. It would be very interesting to repeat this exercise for the quantities we have identified; we hope to return to this in the near future. Ultimately, it would be fascinating to find connections between the sigma-model describing random matrix statistics, and the sigma-models describing other properties of chaotic CFTs such as out-of-time-order correlators \cite{Haehl:2018izb} or even hydrodynamics \cite{Haehl:2018lcu,Liu:2018kfw,Winer:2020gdp}.

An important caveat in our analysis was the restriction to the continuous part of the $SL(2,\mathbb{Z})$ spectrum. For a ramp at spin 0 this restriction is irrelevant. However, also for higher spin sectors our analysis is expected to hold for the discrete Maass cusp forms in a similar way, though the argument is more numerical \cite{toappear}. Nevertheless we can already draw the following conclusion from studying just the Eisenstein series: if the universal ramp is encoded {\it at least partially}\footnote{By ``partially encoded'' we mean that the coefficient of the ramp in the Eisenstein sector is not zero.} in the correlations of Eisenstein series overlap coefficients of a particular spin sector, then spectral determinacy is in general not sufficient to conclude random matrix universality in the other spin sectors. Our conclusion would only fail to hold in cases where the ramp is encoded {\it solely} in the sector of Maass cusp forms, as we didn't analyze this sector here.\footnote{Thus, we merely make an assumption of genericity, namely that there do not exist any constraints that prevent the ramp from being encoded in the continuous spectrum.} It would be interesting to study if there exist general criteria that force the ramp for $m>0$ to be encoded in either the continuous or the discrete sectors alone.

With a view towards other future applications, we note that the formula expressing spectral determinacy \eqref{eq:calZcalZtrf}, as a precise and explicit rephrasing of the constraints of modular invariance, may be useful in implementations of the modular bootstrap. 

We used as a starting point the results from the gravitational analysis \cite{Cotler:2020hgz,Cotler:2020ugk} regarding the existence of random matrix statistics in each spin sector. However, one may hope to {\it derive} such statistics using the modular and conformal bootstraps, or show that such statistics is consistent with the universal properties of conformal field theories.\footnote{One might analyze the accumulation of operators near extremality in CFTs with a twist gap, c.f., \cite{Pal:2022vqc}.} We have shown that such statistics is consistent with modular invariance, and thus have taken the first steps towards such a proof.

In our analysis of the modular S-transform of the spectral form factor, an extension of the saddle point analysis in section \ref{sec:sdual-ramp} outside of the decompactification limit would strengthen our claims regarding the extended regime of validity of random matrix universality. It would be interesting to develop more rigorous techniques to analyze this (e.g., along the lines of \cite{Mukhametzhanov:2019pzy,Pal:2019zzr}).

\vspace{10pt}
\section*{Acknowledgments}

We thank Eric Perlmutter for comments on a draft and Scott Collier for very helpful comments on the role of Maass cusp forms. We further thank Jonah Berean-Dutcher, Kristan Jensen, Henry Maxfield, Chistopher Waddell, David Wakeham for useful conversation. FH is supported by the UKRI Frontier Research Grant EP/X030334/1 and acknowledges support by the DOE grant DE-SC0009988 during the early stages of this project. CM, WR and MR are supported by a Discovery grant from NSERC.

\vspace{10pt}
\appendix

\vspace{5pt}
\section{Notation and conventions}
\label{app:notation}

In this appendix we collect some definitions and conventions used in the main text. 

We use the following inner product for square-integrable functions on the fundamental domain ${\cal F} \equiv \mathbb{H}/SL(2,\mathbb{Z})$:
\begin{equation}
  ( f,g) = \int_{\cal F} \frac{dx dy}{y^2} \, f(x+iy) \, \overline{g(x+iy)} \,.
\end{equation} 
The real-analytic Eisenstein series $E_{\frac{1}{2}+i\alpha}$ and the (normalized) Maass cusp forms $\nu_n$ furnish orthogonal bases of the continuous and discrete eigenfunctions of the Laplactian $\Delta_{_{\cal F}} = -y^2(\partial_x^2 + \partial_y^2)$:
\begin{equation}
\begin{split}
E_{s}(\tau=x+iy) &= \left[ y^{s} + \frac{\Lambda(1-s)}{\Lambda(s)} \, y^{1-s} \right] + \sum_{m\geq 1} \cos(2\pi m x)\, \frac{4\,\sigma_{2s-1}(m)}{m^{s-\frac{1}{2}}\Lambda\left( s\right)} \, \sqrt{y} K_{s-\frac{1}{2}}(2\pi m y)\,,\\
\nu_n(\tau=x+iy) &= \sum_{m\geq 1}\cos(2\pi  m x)\, a_m^{(n)} \, \sqrt{y} K_{iR_n} (2\pi m y) \,,
\end{split}
\end{equation}
where we remind the reader that we only consider even (in $x$) Maass cusp forms in this paper. These functions satisfy 
\begin{equation}
    \Delta_{_{\cal F}} E_s(\tau) = s(1-s) E_s(\tau) \,,\qquad \Delta_{_{\cal F}} \nu_n(\tau) = \left( \frac{1}{4} + R_n^2 \right) \nu_n(\tau) \,.
\end{equation}
For any normalizable modular invariant function $f\in L^2({\cal F})$, we have the Roelke-Selberg spectral decomposition:
\begin{equation}
 f(\tau) = \frac{1}{4\pi} \int_{-\infty}^\infty d\alpha \, \big( f,\, E_{\frac{1}{2}+i\alpha} \big) \, E_{\frac{1}{2}+i\alpha}(\tau) + \sum_{n\geq 0} \frac{(f,\,\nu_n)}{(\nu_n,\nu_n)} \, \nu_n(\tau) \,. 
\end{equation}
The unfolding trick allows one to reduce the inner product with Eisenstein series to an integral over their spin 0 part:
\begin{equation}
  (f,E_{\frac{1}{2}+i\alpha}) = \int_{\cal F} \frac{dxdy}{y^2} \, f(x+iy) E_{\frac{1}{2}-i\alpha}(x-iy) = \int_0^\infty dy \, y^{-\frac{3}{2} - i \alpha} \, f^{m=0}(y)
\end{equation}
where $f^{m=0}(y) \equiv \int_{-\frac{1}{2}}^{\frac{1}{2}} dx \, f(x+iy)$ is the spin 0 component of $f$.\\

We frequently use modified Bessel functions. These can be written as 
\begin{equation}
    K_{i\alpha} (2\pi m y) = \frac{1}{2}\int_{-\infty}^\infty d\xi \; e^{-2\pi m y \cosh \xi} \, \cos(\alpha \xi) \label{eq:xi-definitionApp}
\end{equation}
and they satisfy the following orthogonality relation, found in \cite{Bielski}:
\begin{equation}
    \int_0^{\infty} \frac{dy}{y} \, K_{i \alpha}(2\pi m y) K_{i\alpha'}(2\pi m y) = \frac{\pi^2}{2 \alpha \sinh(\pi \alpha)}\,\left[ \delta(\alpha-\alpha') + \delta(\alpha+\alpha') \right]\,,\label{eq:Kdirac}
\end{equation}
as well as the identities \cite{gradshteyn2007}
\begin{equation}
\begin{split}
 \frac{1}{\pi}\int_{-\infty}^\infty d\alpha \, \alpha \, \tanh(\pi \alpha) \, K_{i\alpha}(2\pi m_1 y_1) K_{i\alpha}(2\pi m_2y_2) &=  \frac{\sqrt{m_1y_1\, m_2y_2}}{m_1y_1+m_2y_2} \, e^{-2\pi (m_1y_1+m_2y_2)} \,,\\
  \int_0^\infty \frac{dy_1}{\sqrt{y_1}} \, \frac{e^{-2\pi m_1y_1}}{ y_1+y_2} \, K_{i\alpha}(2\pi m_1 y_1) &= \frac{\pi\, e^{2\pi m_1 y_2}}{\sqrt{y_2} \, \cosh(\pi \alpha)}\, K_{i\alpha}(2\pi m_1 y_2) \,.
 \end{split}
 \label{eq:Kdirac2}
\end{equation}

\vspace{5pt}
\section{Change of bases and the Maass cusp forms}
\label{app:maass}

For completeness, we show here the equivalent of the expressions given in section \ref{sec:xibasis} for the discrete part of the spectrum. The discrete part of the partition function in $\xi$ variables is
\begin{equation}
\begin{split}
    \widetilde{Z}^m_\text{P,disc.}(y) &= \sum_{n} z_n \, a_m^{(n)} \, \sqrt{y} K_{iR_n}(2\pi m y) \\
    &=\frac{\sqrt{y}}{2\pi} \int d\xi \,{\cal Z}^m_\text{P,disc.}(\xi) \, e^{-2\pi m y \cosh \xi}
\end{split}
\end{equation}
where 
\begin{equation}
    {\cal Z}^m_\text{P,disc.}(\xi) \equiv \pi \sum_{n} \,\cos(R_n \xi)\,  a_m^{(n)} \, \,z_{n} \,.
\end{equation}
Inverting this, the $z_n$ are given by
\begin{equation}
    \sum_n a_m^{(n)} \, z_n \, [\delta(R-R_n)+\delta(R+R_n)] =\frac{1}{\pi^2}\int d\xi \, \cos(R\xi) \, {\cal Z}^m_\text{P,disc.}(\xi) 
\end{equation}
This expression is more complicated to work with than the analogous one for Eisenstein series. It is natural to conjecture that we can still express the discrete spin $m$ partition function (in $\xi$-space) through the spin $m'$ partition function via a particular kernel:
\begin{equation}
\label{eq:SpecDetMaass}
    {\cal Z}^{m}_\text{P,disc.}(\xi) \stackrel{?}{=}  \int_{-\infty}^\infty d\xi' \, {\cal K}^{m,m'}_\text{disc.}(\xi,\xi') \, {\cal Z}_\text{P,disc.}^{m'}(\xi')
\end{equation}
However, the precise form of the kernel ${\cal K}^{m,m'}$ is not simple and requires regularization and a numerical construction. In any case, it will depend on the eigenvalues $R_n$ and the Fourier coefficients $a_m^{(n)}$ in a nonlocal way and it will require the Fourier coefficients to be non-degenerate and non-vanishing; this is a well-known unproven conjecture, but it is most likely true \cite{hejhal92}.

We also note that the kernels ${\cal K}^{m,m'}$ (for both the continuous and discrete parts) have some nice properties: since $\frac{\sigma_{2i\alpha}(m)}{m^{i\alpha}}$ and $a_m^{(n)}$ are eigenvalues of Hecke operators, they satisfy:
\begin{equation}
    \left( \frac{\sigma_{2i\alpha}(m)}{m^{i\alpha}} \right) \left( \frac{\sigma_{2i\alpha}(m')}{m'^{\,i\alpha}} \right) = \sum_{\substack{ \ell | (m,m') \\ \ell > 0 }} \left( \frac{\sigma_{2i\alpha}\big(\frac{mm'}{\ell^2}\big)}{\big(\frac{mm'}{\ell^2}\big)^{i\alpha}}\right) \,,\qquad 
    a_m^{(n)} a_{m'}^{(n)} = \sum_{\substack{ \ell | (m,m') \\ \ell > 0 }} a^{(n)}_{\frac{mm'}{\ell^2}}
\end{equation}
In particular, the coefficients `multiply' for prime spins:
\begin{equation}
    \sigma_{2i\alpha}(p) \,\sigma_{2i\alpha}(p') = \sigma_{2i\alpha}(pp') 
    \quad \text{ for } p,p' \text{ prime.}
\end{equation}
Hence, the only independent data about the kernels consists of $\frac{\sigma_{2i\alpha}(m)}{m^{i\alpha}}$ and $a_m^{(n)}$ for prime $m$. For any other values of $m$, these coefficients are determined as linear combinations of those for prime $m$.

\section{Numerical data}
\label{app:numerics}

In this appendix we collect some numerical results, complementing the discussion in sections \ref{sec:spin0imprint} and \ref{sec:spin1imprint}.

\begin{figure}
    \centering
\includegraphics[width=.8\textwidth]{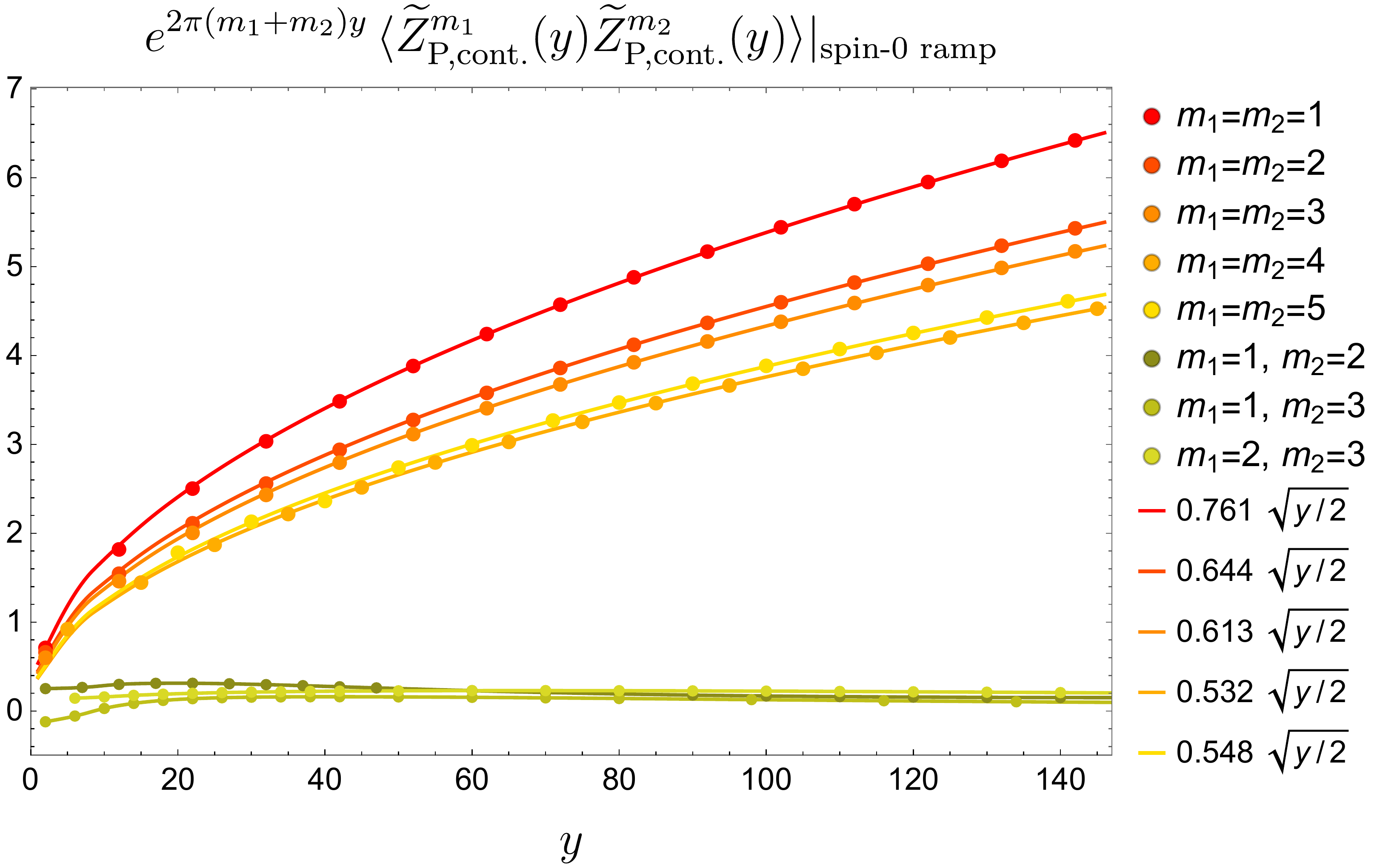}
    \caption{The result of numerical evaluation of \eqref{eq:ZmZmInt}: the contribution to the spin $(m_1,m_2)$ spectral form factor that originates from a spin 0 ramp through symmetries. We plot as a function of $y_1=y_2 \equiv y$. The plot confirms that this does {\it not} take the form of a ramp, but is rather a subleading correction to it, \eqref{eq:ZZresNum}. Universal eigenvalue repulsion in the spin 0 sector hence doesn't imply the same in any other spin sector through symmetries (despite spectral determinacy in the exact partition function).}
    \label{fig:ZmZm1}
\end{figure}

\begin{figure}
    \centering
\includegraphics[width=.8\textwidth]{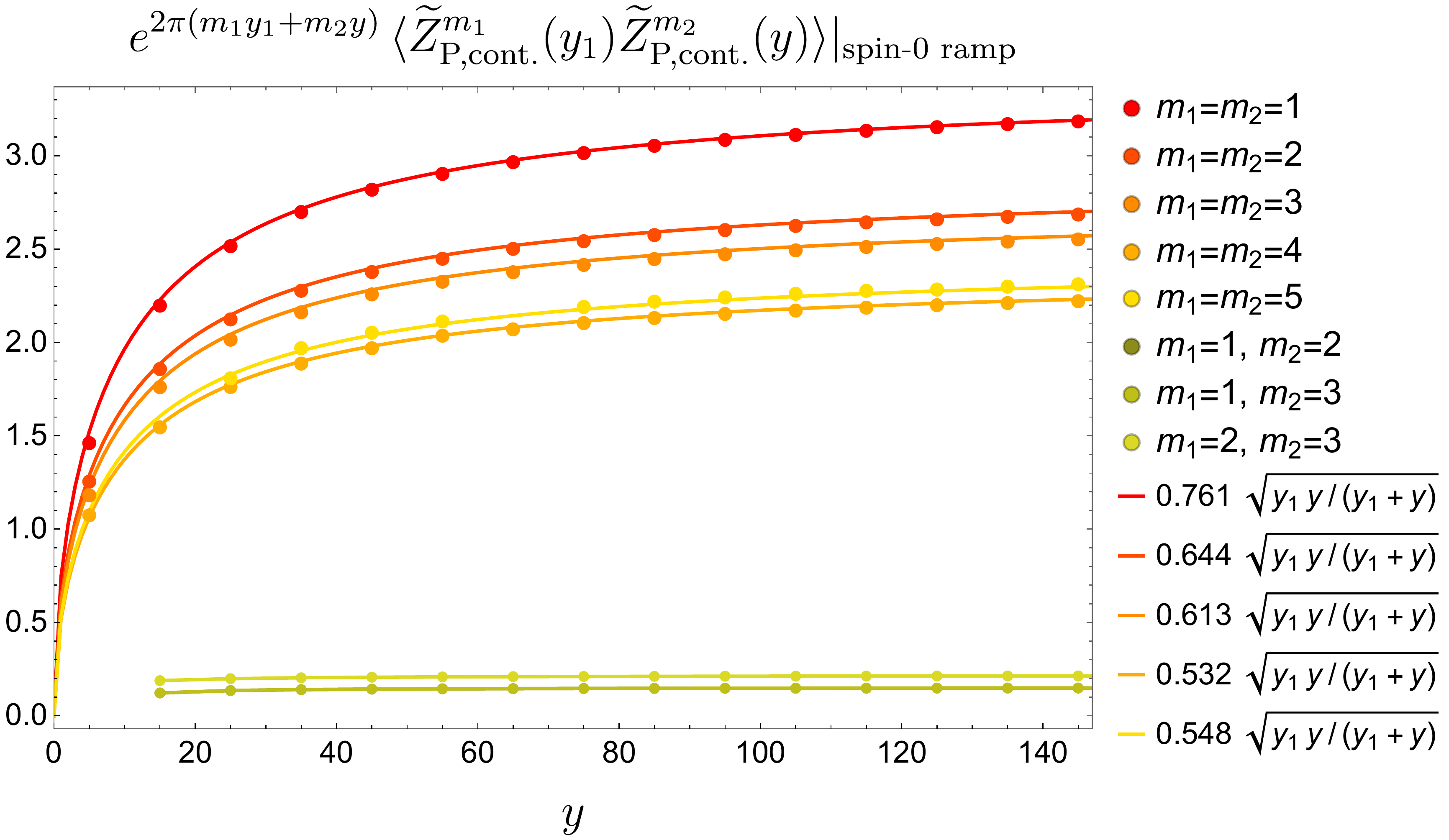}
    \caption{The result of numerical evaluation of \eqref{eq:ZmZmInt}, c.f., figure \ref{fig:ZmZm1}. In this plot we fix $y_1=20$ and plot as a function of $y_2 \equiv y$. Again, the result is well fitted by \eqref{eq:ZZresNum}.}
    \label{fig:ZmZm2}
\end{figure}

Figures \ref{fig:ZmZm1} and \ref{fig:ZmZm2} show the result \eqref{eq:ZZresNum} for different cross sections of the $(y_1,y_2)$-plane. For example, we show this behavior and the numerical data points for $y_1 = y_2 \equiv y$ in figure \ref{fig:ZmZm1}. The case of unequal spins yields curves, which are consistent with contributions that are strongly suppressed compared to the equal spin spectral correlators, hence supporting the proportionality to $\delta_{m_1m_2}$.
Similarly, figure \ref{fig:ZmZm2} shows the same analysis for $y_1=20$ as a function of $y_2 \equiv y$. It is analogous to figure \ref{fig:ZmZm1}, but shows a different section of the $(y_1,y_2)$-plane.
We have also analyzed other cross sections of the $(y_1,y_2)$-plane in a similar manner, in order to ascertain the functional form of \eqref{eq:ZZresNum}.


\begin{figure}
    \centering
\includegraphics[width=.8\textwidth]{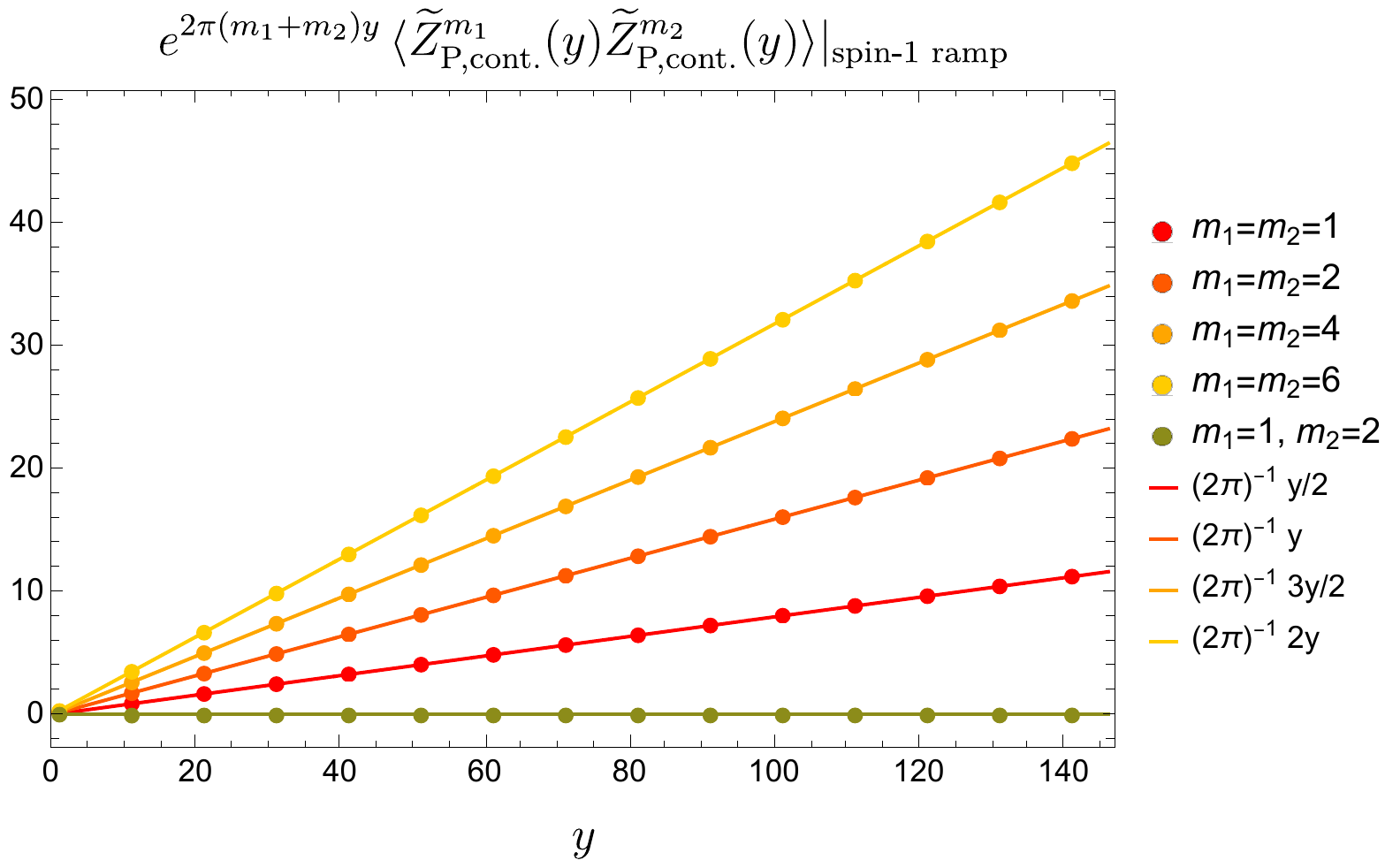}
    \caption{The result of numerical evaluation of \eqref{eq:ZmZmspin1Imprint}, i.e., the imprint of a ramp at spin 1 onto the spin $(m_1,m_2)$ spectrum, together with a linear fit (solid lines). The $y$-scaling is linear (as expected for a ramp), but the coefficient doesn't match the random matrix expectation for any $m_1=m_2>1$.}
    \label{fig:ZmZm3}
\end{figure}

Similarly, in figure \ref{fig:ZmZm3} we show the result of numerically evaluating \eqref{eq:ZmZmspin1Imprint}, again as a function of $y_1=y_2 \equiv y$; other cross sections of the $(y_1,y_2)$ plane can be checked similarly. We performed the evaluation by assuming a saddle point approximation, i.e., approximating the exponent in \eqref{eq:ZmZmspin1Imprint} by its second order expansion in small $\xi_i$. The plot confirms the result \eqref{eq:Zm1Zm1result}: we find a functional form that is actually consistent with a ramp, but an overall coefficient that is not. This information is thus not sufficient to conclude random matrix universality at spin $(m_1,m_2)$. 
\newpage

\section{Details on the S-transform by saddle point}
\label{sec:saddle-details}

Here we provide details on the saddle point calculation of section \ref{sec:sdual-ramp}, and also discuss subleading corrections. 

\subsection{Saddle point evaluation}

Consider the Fourier transforms \eqref{eq:saddleCalc0} and focus for clarity on the first set of variables $(P_1',\bar{P}_1')$. The Fourier transform in only these variables is given by
{\small
\begin{equation}
\label{eq:saddleCalc1}
\begin{split}
&\langle \rho_\text{P}(P_1,\bar{P}_1)\rho_\text{P}(P_2',\bar{P}_2')\rangle \\
&\quad\approx  2\int dP_1' d\bar{P}_1'\,  e^{-4\pi i (P_1' P_1+\bar{P}_1'\bar{P}_1)}\sum_{m_i\in \mathbb{Z}} F^{m_1,m_2}(P_1',\bar{P}_1',P_2',\bar{P}_2') \,\delta(P_1'^2-\bar{P}_1'^2-m_1)\delta(P_2'^2-\bar{P}_2'^2-m_2)\\
&\quad= 4\sum_{m_i\in \mathbb{Z}}  \int d\lambda_1\,e^{4\pi im_1\lambda_1}\left( \int dP_1' d\bar{P}_1' \, e^{-4\pi i (P_1' P_1+\bar{P}_1'\bar{P}_1+\lambda_1(P_1'^2 -\bP_1'^2))}F^{m_1,m_2}(P_1',\bar{P}_1',P_2',\bar{P}_2')\right)\delta(P_2'^2-\bar{P}_2'^2-m_2)
\end{split}
\end{equation}
}\normalsize
where we exponentiated the delta-function.\footnote{In the case $m_i'=0$, one can simply use the Dirac delta functions without exponentiating them. This will end up being subleading in our large parameters.} We now consider $P_i^2,\bP_i^2\gg 1$ and find the saddle point of the oscillatory exponential (the function $F^{m_1,m_2}$ is slowly varying, hence amenable to saddle point analysis; as pointed out in footnote \ref{foot:reg}, we assume that the singularity as $\omega \rightarrow 0$ is regulated in a more complete description). The saddle points are $(P_{1*}',\bP_{1*}')=\big(\! -\!\frac{P_1}{2\lambda_1},\frac{\bP_1}{2\lambda_1}\big)$. The result is
\begin{equation}
{\small
\begin{aligned}
&\langle \rho_\text{P}(P_1,\bar{P}_1)\rho_\text{P}(P_2',\bar{P}_2')\rangle\\
&\;\; \approx \sum_{m_i\in \mathbb{Z}} \int \frac{ d\lambda_1}{|\lambda_1|} \; e^{4\pi i\left(m_1\lambda_1+\frac{P_1^2-\bar{P}_1^2}{4\lambda_1}\right)} \,F^{m_1,m_2}\left(-\frac{P_1}{2\lambda_1},\frac{\bar{P}_1}{2\lambda_1},P_2',\bar{P}_2'\right)\delta(P_2'^2-\bar{P}_2'^2-m_2)
\qquad (P_1^2,\bP_1^2 \gg 1)
\end{aligned}
}\normalsize
\label{eq:saddle-first}
\end{equation}
We can now evaluate the integral over $\lambda_1$ by another saddle point approximation. The large parameter is $\sqrt{m_1(P_1^2-\bP_1^2)}\equiv\sqrt{m_1\mc_1}$, and the saddle points are $\lambda_{1*}= \pm \frac{1}{2}\sqrt{\frac{\mc_1}{m_1}}$; this is real when $m_1,\mc_1$ have the same sign, and imaginary if they have opposite sign. In the latter case, one can show that the integral decays exponentially, and thus can be neglected. 

After doing the integrals over $P_2',\bP_2'$ the same way, we find:
\begin{equation}
\begin{split}
     &\langle \rho_\text{P}(E_1,\mc_1) \rho_\text{P}(E_2,\mc_2)\rangle \\
     &\qquad \approx 2 \sum_{\substack{m_i\in \mathbb{Z}:\\m_i \mc_i>0}}  F^{m_1,m_2}\left(P_1\sqrt{\frac{m_1}{\mc_1}},\bP_1\sqrt{\frac{m_1}{\mc_1}},P_2\sqrt{\frac{m_2}{\mc_2}},\bP_2\sqrt{\frac{m_2}{\mc_2}}\right)\,|\mc_1 \mc_2 m_1 m_2|^{-1/4} \\
     &\qquad\qquad\qquad\qquad\quad \times \cos\left(4\pi\sqrt{m_1 \mc_1}+\frac{\pi}{4}\right)\cos\left(4\pi\sqrt{m_2 \mc_2}+\frac{\pi}{4}\right) \qquad \quad (P_i^2 , \bP_i^2, |\mc_i| \gg 1)\,.
\end{split}
\label{eq:s-dual-saddle-prerampGeneral}
\end{equation}
Assuming for cleaner notation $\mc_i>0$, we obtain
\eqref{eq:s-dual-saddle-preramp}.

\subsection{Corrections}

Because the $P_i', \, \bP_i',$ and $\lambda_i$ saddle points depend on different large parameters, there is a possibility that subleading terms in one saddle point calculation (for example, the $\bP_i$ saddle point) can actually dominate in another saddle point analysis (for example, the $\lambda_i$ saddle point). We will show that this is not the case, thus providing a consistency check. We assume throughout that $P_i\gg\bP_i$.

First, we consider the subleading terms for the $P_1', \, \bP_1',\, P_2',\, \bP_2'$ saddle points. We use the method of steepest descent to put the integrals in a form amenable to subleading analysis. Focusing on the $P_1', \, \bP_1'$ integrals for clarity, we first do a change of variables $P_1'\rightarrow \frac{P_1}{2\lambda_1} P_1', \, \bP_1'\rightarrow \frac{\bP_1}{2\lambda_1} \bP_1'$ which turns \eqref{eq:saddleCalc1} into
\begin{equation}
{\small
\begin{aligned}
&\langle \rho_\text{P}(P_1,\bar{P}_1)\rho_\text{P}(P_2',\bar{P}_2')\rangle = \sum_{m_i\in \mathbb{Z}}  \int d\lambda_1\,e^{4\pi im_1\lambda_1}\,\delta(P_2'^2-\bar{P}_2'^2-m_2) \\
&\qquad \times \int dP_1' d\bar{P}_1' \, \frac{P_1 \bP_1}{\lambda_1^2} \, e^{2\pi \frac{P_1^2}{|\lambda_1|}\left(-i\sgn(\lambda_1)(P_1'+\frac{1}{2}P_1'^2)\right) + 2\pi \frac{\bP_1^2}{|\lambda_1|}\left(-i\sgn(\lambda_1)(\bP_1'-\frac{1}{2}\bP_1'^2)\right)   }F^{m_1,m_2}\left(\frac{P_1}{2\lambda}P_1',\frac{\bP_1}{2\lambda}\bar{P}_1',P_2',\bar{P}_2'\right)
\end{aligned}
}\normalsize
\end{equation}
Our large, positive parameters are $2\pi \frac{P_1^2}{|\lambda_1|}, \, 2\pi \frac{\bP_1^2}{|\lambda_1|}$. We rotate our integration contours so that the exponent has constant imaginary part, passes through the saddle point, and is exponentially decaying. The contours that achieve this are parametrized as
\begin{equation}
    P_1'(u_1)=u_1-i\sgn(\lambda_1)(u_1+1)\, , \quad \bP_1'(\bar{u}_1)=\bar{u}_1+i\sgn(\lambda_1)(\bar{u}_1-1)\,,
\end{equation}
which gives 
\begin{equation}
{\small
\begin{aligned}
&\langle \rho_\text{P}(P_1,\bar{P}_1)\rho_\text{P}(P_2',\bar{P}_2')\rangle = \sum_{m_i\in \mathbb{Z}}  \int d\lambda_1\;e^{4\pi i\left(m_1\lambda_1+\frac{P_1^2-\bP_1^2}{4\lambda_1}\right)} \,\frac{2P_1 \bP_1}{\lambda_1^2}  \,
 \delta(P_2'^2-\bar{P}_2'^2-m_2) \\
&\times\int d^2u_1 \, e^{-2\pi \frac{P_1^2}{|\lambda_1|}(u_1+1)^2 - 2\pi \frac{\bP_1^2}{|\lambda_1|}(\bar{u}_1-1)^2}F^{m_1,m_2}\left(\frac{P_1}{2\lambda_1}\left(u_1-i\sgn(\lambda_1)(u_1+1)\right),\frac{\bP_1}{2\lambda_1}\left(\bar{u}_1+i\sgn(\lambda_1)(\bar{u}_1-1)\right),P_2',\bar{P}_2'\right)
\end{aligned}
}\normalsize
\end{equation}
The corrections to the leading result all come from the Taylor series of $F^{m_1,m_2}(\cdots)$ at $u_1=-1, \, \bar{u}_1=1$. We apply the same procedure to $P_2',\, \bP_2'$ and get the general form of the corrections to the leading saddle:
\begin{equation}
{\small
\begin{aligned}
    &\text{corrections} \propto \frac{\lambda_1^{n_1 + \bar{n}_1} \lambda_2^{n_2+\bar{n}_2}}{P_1^{2n_1}\bP_1^{2\bar{n}_1}P_2^{2n_2}\bP_2^{2\bar{n}_2}}\partial^{2n_1}_{u_1} \partial^{2\bar{n}_1}_{\bar{u}_1} \partial^{2n_2}_{u_2} \partial^{2\bar{n}_2}_{\bar{u}_2}
    \\
    &\quad\times  F^{m_1,m_2}\left(\frac{P_1}{2\lambda_1}\left(u_1-i(u_1+1)\right),\frac{\bP_1}{2\lambda_1}\left(\bar{u}_1+i(\bar{u}_1-1)\right),\frac{P_2}{2\lambda_2}\left(u_2-i(u_2+1)\right),\frac{\bP_2}{2\lambda_2}\left(\bar{u}_2+i(\bar{u}_2-1)\right)\right)
\end{aligned}
}\normalsize
\end{equation}
where we have factored out the universal part of the corrections and chosen $\lambda_i>0$ for convenience. The potential issue is now clear; the prefactor contains positive powers of $\lambda_i$, whose saddle point is proportional to $\sqrt{\mc_i} \approx P_i$, and the function contains non-trivial dependence on $\lambda_i$. However, explicitly checking this issue using \eqref{eq:ramp-p-vars} reveals that these corrections are indeed subleading with respect to our large parameters. 

Additionally, our saddle point parameter for the $\bP_i'$ integral, $2\pi\frac{\bP_i}{2|\lambda|} \approx \pi \frac{\bP_i^2}{P_i}$, is not guaranteed to be large for all possible regimes of $P_i^2\gg \bP_i^2\gg1$. While at first this suggests a more rigorous scaling is required, explicit analysis of the corrections reveals that any possible relative scaling of $P_i^2, \, \bP_i^2$ still leads to a subleading correction. 

Knowing now that keeping just the leading term is consistent, we analyze the subleading corrections to the $\lambda_i$ saddle point. The exact form of these corrections is significantly more complicated, as the exponent in the integral is not Gaussian. However, we can still analyze how these corrections will depend on our large parameters. We perform a change of variables $\lambda_i \rightarrow \frac{1}{2}\sqrt{\frac{\mc_i}{m_i}}\,\lambda_i$, and have 
\begin{equation}
\begin{split}
&\langle \rho_\text{P}(P_1,\bar{P}_1)\rho_\text{P}(P_2,\bar{P}_2)\rangle\\
&\quad \approx \sum_{m_i\in \mathbb{Z}} \int \frac{ d\lambda_1}{|\lambda_1|}\frac{ d\lambda_2}{|\lambda_2|} \; e^{2\pi\sqrt{\mc_1 m_1} i\left(\lambda_1+\frac{1}{\lambda_1}\right)+2\pi\sqrt{\mc_2 m_2} i\left(\lambda_2+\frac{1}{\lambda_2}\right)} 
\\
&\qquad \qquad\qquad\times F^{m_1,m_2}\left(-P_1\sqrt{\frac{m_1}{\mc_1}}\,\lambda_1^{-1},\bP_1\sqrt{\frac{m_1}{\mc_1}}\,\lambda_1^{-1},-P_2\sqrt{\frac{m_2}{\mc_2}}\,\lambda_2^{-1},\bP_2\sqrt{\frac{m_2}{\mc_2}}\,\lambda_2^{-1})\right)
\end{split}
\end{equation}
The large parameters are $2\pi \sqrt{m_i \mc_i}$, and the saddle points are at $\lambda_i=\pm 1$. However the steepest descent contour $\lambda_1(t)=\lambda_2(t)\equiv\lambda(t)$ is quite complicated. Additionally, subleading terms do not come just from a Taylor expansion of $F^{m_1,m_2}(\cdots)$, but also from expanding the exponent $\phi(t)=i\left(\lambda(t_i)+\frac{1}{\lambda(t_i)}\right)$ beyond quadratic order. Thankfully, all the dependence on the large parameters comes only from $(i)$ powers of $\sqrt{m_i \mc_i}$, and $(ii)$ the dependence of the function $F^{m1,m2}$ on the large parameters, which is insensitive to the details of the steepest descent contours. Thus, the corrections are of the general form
\begin{equation}
{\small
\begin{aligned}
    &\text{corrections}\propto \, \, (m_1 \mc_1)^{-n_1/2} (m_2 \mc_2)^{-n_2/2}\, \partial^{c_1}_{t_1}\partial^{c_2}_{t_2}
    \\
    & \qquad\qquad\qquad\quad\times F^{m_1,m_2}\left(-P_1\sqrt{\frac{m_1}{\mc_1}}\lambda(t_1)^{-1},\bP_1\sqrt{\frac{m_1}{\mc_1}}\lambda(t_1)^{-1},-P_2\sqrt{\frac{m_2}{\mc_2}}\lambda(t_2)^{-1},\bP_2\sqrt{\frac{m_2}{\mc_2}}\lambda(t_2)^{-1})\right)
\end{aligned}
}\normalsize
\end{equation}
with $c_i \leq 2n_i$. Here, the lower order derivative terms come from terms where the exponential is expanded.\footnote{For a standard one-dimensional Laplace integral, $\int dx \, e^{a \phi(x)} f(x)$, there are higer order terms such as $\frac{1}{a}\frac{f'(c)\phi'''(c)}{2\phi''(c)^2}$. } Explicitly checking using \eqref{eq:ramp-p-vars}, we see that all corrections are indeed subleading with respect to all our large parameters, and they do not depend on the specifics of the relative scaling of $P_i^2, \, \bP_i^2$.

We also briefly mention the comment after \eqref{eq:s-dual-saddle-preramp}. Strictly speaking, for $\mc_i$ large but fixed, because we sum over all spins there will eventually be a regime where $\sqrt{\frac{m_i}{\mc_i}}$ is not small. We assume that these asymptotic contributions are negligible; however this does require us to assume we can exchange the limits $\mc_i\rightarrow \infty$ and $M\rightarrow \infty$. That we get a sensible result in the decompactification limit suggests this is a reasonable assumption; addressing this in detail with more care would require an analysis of the full modular invariant density of states, which is beyond the scope of this paper.

\vspace{10pt}
\bibliographystyle{JHEP}
\newpage

\providecommand{\href}[2]{#2}\begingroup\raggedright\endgroup

\end{document}